\newcommand{\CLASSINPUTbottomtextmargin}{2cm}

\documentclass[journal,draftclsnofoot,onecolumn,12pt]{IEEEtran}

\usepackage{amsthm,amssymb,graphicx,multirow,amsmath,color,amsfonts}
\usepackage[update,prepend]{epstopdf}
\usepackage[noadjust]{cite}
\usepackage[latin1]{inputenc}
\usepackage{pdfpages}
\usepackage{comment}
\usepackage{hyperref}
\usepackage{subcaption}
\usepackage{csquotes}
\usepackage[table]{xcolor}
\definecolor{lightgray}{gray}{0.9}
\usepackage{todonotes}

\usepackage{setspace}	

\allowdisplaybreaks 
\def\matern{{Mat{\'e}rn\;}}

\def\caseS{{\sc case}}
\def\caseC{{\sc Case}}

\def\h{\ensuremath{\bar{f}}}
\newcommand{\myeq}[1]{\mathrel{\overset{\makebox[0.07pt]{\mbox{(#1)}}}{=}}}
\newcommand{\myeqq}[1]{\mathrel{\overset{\makebox[0.01pt]{\mbox{$#1$}}}{=}}}

\def\nb0{{\mathbf{0}}}
\def\nb1{{\mathbf{1}}}

\def\ncalB{{\mathcal{B}}}

\def\ncalH{{\mathcal{H}}}
\def\ncalI{{\mathcal{I}}}

\def\ncalK{{\mathcal{K}}}

\def\ncalN{{\mathcal{N}}}

\def\ncalR{{\mathcal{R}}}

\def\nbbE{{\mathbb{E}}}

\def\nbbN{{\mathbb{N}}}

\def\nbbP{{\mathbb{P}}}

\def\nbbR{{\mathbb{R}}}

\def\nbbZ{{\mathbb{Z}}}

\newtheorem{lemma}{Lemma}

\newtheorem{ndef}{Definition}

\newtheorem{theorem}{Theorem}
\newtheorem{prop}{Proposition}
\newtheorem{cor}{Corollary}

\newtheorem{remark}{Remark}

\def\figref#1{Fig.\,\ref{#1}}%

\def\pc{\mathtt{P_c}}

\def\R{\mathbb{R}}

\def\sir{\mathtt{SIR}}

\graphicspath{{./Fig/}}
\title{{3GPP-inspired HetNet Model using Poisson Cluster Process: Sum-product Functionals and Downlink Coverage}
}
\author{Chiranjib Saha, Mehrnaz Afshang, and Harpreet S. Dhillon
\thanks{The authors are with Wireless@VT, Department of ECE, Virginia Tech, Blacksburg, VA, USA. Email: \{csaha,  mehrnaz,  hdhillon\}@vt.edu. The first two authors have contributed equally  to the paper. The support of the US National Science Foundation (Grants CCF-1464293 and CNS-1617896) is gratefully acknowledged. 
A part of this paper was presented at the 2017 Information Theory and Applications (ITA) Workshop in San Diego, CA \cite{saha2017poisson}.
\hfill
Manuscript last updated: \today.
} }

\let\emptyset\varnothing
\begin{document}

\maketitle
\thispagestyle{empty}
\pagestyle{empty}
\vspace{-1.2cm}

\begin{abstract}
The growing complexity of heterogeneous cellular networks (HetNets) has necessitated a variety of user and base station (BS) configurations to be considered for realistic performance evaluation and system design. This is directly reflected in the HetNet simulation models proposed by standardization bodies, such as the third generation partnership project (3GPP). Complementary to these simulation models, stochastic geometry-based approach, modeling the locations of the users and the $K$ tiers of BSs as independent and homogeneous Poisson point processes (PPPs), has gained prominence in the past few years. Despite its success in revealing useful insights, this PPP-based $K$-tier HetNet model is not rich enough to capture spatial coupling between user and BS locations that exists in real-world HetNet deployments and is included in 3GPP simulation models.  
In this paper, we demonstrate that modeling a fraction of users and arbitrary number of BS tiers alternatively with a Poisson cluster process (PCP) captures the aforementioned coupling, thus bridging the gap between the 3GPP simulation models and the PPP-based analytic model for HetNets. We further show that the downlink coverage probability of a typical user under maximum signal-to-interference-ratio ($\sir$) association can be expressed in terms of the {\em sum-product functionals} over PPP, PCP, and its associated offspring point process, which are all characterized as a part of our analysis.
 We also  show that the proposed model converges to the PPP-based HetNet model as the cluster size of the PCPs tends to  infinity. Finally,
 we  specialize our analysis based on general PCPs  for   Thomas and \matern cluster processes.  
Special instances  of the proposed  model closely resemble the different configurations for BS and user locations considered in 3GPP simulations.%
\end{abstract}  
\vspace{-0.9em}

\section{Introduction}
In order to handle the exponential growth of mobile data traffic, macrocellular networks of  yesteryears have gradually evolved into more denser heterogeneous cellular networks in which several types of low power BSs (called small cells) coexist with macrocells. While macro BSs (MBSs) were deployed fairly uniformly to provide a ubiquitous coverage blanket, the small cell BSs (SBSs) are deployed somewhat organically to complement capacity of the cellular networks (primarily at user hotspots) or to patch their coverage dead-zones.  
This naturally couples the locations of the SBSs with those of the users, as a result of which we now need to consider plethora of deployment scenarios in the system design phase as opposed to only a few in the macro-only networks of the past. While the simulation models considered by 3GPP are cognizant of this evolution and consider several different configurations of user and SBS locations~\cite{3gppreportr12,3gppreportr13}, the stochastic geometry-based analyses of HetNets still rely on the classical PPP-based $K$-tier HetNet model~\cite{dhillon2012modeling,5743604}, which is not rich enough to capture aforementioned coupling. 
In this paper, we show that this ever-increasing gap between the PPP-based HetNet model and the real-word deployments can be reduced by modeling a fraction of users and an arbitrary number of BS tiers using PCPs. In order to put this statement and our contribution in context, we summarize the state-of-the-art 3GPP and stochastic geometry-based HetNet models next.

%


\subsection{3GPP Models for HetNets}\label{subsec::3gpp::models}

In this section, we summarize models used for system-level simulations by 3GPP.  
For modeling macrocells, 3GPP simulation scenarios rely on either a single macrocell setup or grid based models, where finite number of MBSs are placed as regularly spaced points on a plane. On the contrary, as discussed next, several different configurations corresponding to a variety of real-life deployment scenarios are considered for modeling the locations of users and SBSs (usually pico and femto cells) \cite[Section~A.2.1.1.2]{access2010further}. Some configurations of interest for this paper are summarized in Table~\ref{table::3gpp_configuration}. In order to be consistent with the 3GPP documents, we will put {\em keywords} reserved for referring to the configurations of users (uniform and clustered) and SBSs (correlated and uncorrelated) in the 3GPP documents in quotation marks. 

\begin{figure}
   \centering
  \begin{subfigure}{.3\textwidth}
  \centering
  \includegraphics[width=\linewidth, page=1]{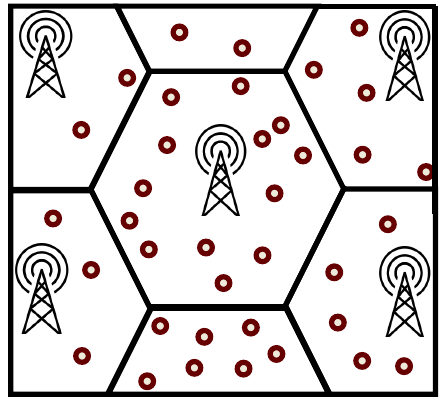}
  \caption{}
  \label{fig:user_uniform_3gpp}
\end{subfigure}%
\hspace{0.2cm}
\begin{subfigure}{.3\textwidth}
  \centering
  \includegraphics[width=\linewidth , page = 2]{./FigN/smallcell2.pdf}
  \caption{}
  \label{fig:user_clustered_3gpp}
\end{subfigure}
\begin{subfigure}{.3\textwidth}
  \centering
  \includegraphics[width=\linewidth , page = 3]{./FigN/smallcell2.pdf}
  \caption{}\label{fig::sbs::3gpp}
  \label{fig:sbs_3gpp}
\end{subfigure}

\caption{\small User and SBS configurations considered in 3GPP HetNet models. Figs. (a) and (b) illustrate two different user configurations: (a) ``uniform''  within a macrocell, and (b) ``clustered'' within a  macrocell. Fig. (c) illustrates SBS configurations: (1) Dense deployment of SBSs at certain areas (usually within user hotspots or indoors), (2) SBSs deployed uniformly at random within a macrocell, and (3) a single SBS deployed within a user hotspot.
 }\label{fig::user_config::3gpp}
\end{figure}

{\em Users.} As illustrated in Figs. \ref{fig:user_uniform_3gpp} and \ref{fig:user_clustered_3gpp}, there are two main user configurations considered in 3GPP simulation models: (i) \enquote{uniform} and (ii) \enquote{clustered}. In the \enquote{uniform} configuration, the users are assumed to be distributed uniformly at random within each macrocell. Given the coverage-centric nature of macrocellular deployments, this configuration has been the default choice for system-level simulations of cellular networks since their inception. However, with the focus quickly shifting towards capacity-driven deployments of SBSs, the \enquote{clustered} user configuration has become at least as much (if not more) important. In this configuration, the users are assumed to be distributed uniformly at random within circular regions of a constant radius (modeling user hotspots). As discussed next, SBSs are often deployed in these user hotspots, which couples their locations with those of the users. 

{\em SBSs.} Roughly speaking, there are two different classes of configurations considered for SBSs: (i) \enquote{uncorrelated} and (ii) \enquote{correlated}. In the \enquote{uncorrelated} configuration, the SBSs are assumed to be distributed uniformly at random inside a macrocell. This corresponds to  configuration 2 in Fig. \ref{fig:sbs_3gpp}. The complete description of \enquote{correlated} configurations is a bit more tedious due to their context-specific nature. Therefore, we will first summarize the factors that introduce correlation or coupling in the SBS locations and then describe the configurations that are most relevant to this paper. {\em Intra-tier} coupling in the SBS locations is introduced when SBSs are deployed according to some site-planning optimization strategies to maximize coverage over the macrocell. {\em Inter-tier} coupling in the SBS and MBS locations is introduced when more SBSs are deployed at the cell-edge to boost cell-edge coverage. Similarly, {\em SBS-user} coupling results from the user-centric deployment of small cells in the user hotspots. Interested readers are advised to refer to \cite{3gppreportr12,3gppreportr13,access2010further} for more details about how these sources of correlation manifest into the 3GPP simulation models. In this paper, we are most interested in the SBS-user coupling. Please refer to \figref{fig::sbs::3gpp} (configurations 1 and 3) for illustrative examples. As will be evident soon, these configurations will appear as special cases of the unified approach proposed in this paper.

\subsection{Stochastic geometry-based approaches}
In parallel to the realistic simulation models used by 3GPP, {\em analytical} HetNet models with foundations in stochastic geometry have gained prominence in the last few years~\cite{elsawy2013stochastic,andrews2016primer,elsawy2016modeling,mukherjee2014analytical}. The main idea here is to endow the locations of the BSs and users with distributions and then use tools from stochastic geometry to derive easy-to-compute expressions for key performance metrics, such as coverage and rate\footnote{A careful reader will note that 3GPP models also endow the locations of users and SBSs with distributions, which technically makes them stochastic models as well.}. In order to maintain tractability, the locations of the users and different types of BSs are usually modeled by independent homogeneous PPPs. We will henceforth refer to homogeneous PPP as a PPP unless stated otherwise. 
 This model,  usually referred to as a $K$-tier HetNet model, was first introduced in~\cite{5743604,dhillon2012modeling} and generalized in several important ways in~\cite{mukherjee2012distribution,jo2012heterogeneous,MadhusudhananRestrepoBrown2016,MIMO6596082}. Reviewing the rich and diverse collection of the followup works is outside the scope of this paper. Interested readers are advised to refer to extensive surveys in \cite{elsawy2013stochastic,andrews2016primer,elsawy2016modeling,mukherjee2014analytical}. Since the fundamental assumption in this PPP-based $K$-tier HetNet model is the mutual independence of all the BS and user locations, it is not rich enough to capture spatial coupling that exists in HetNets. As a result, there have been many attempts in the recent past to use more sophisticated point processes to model different elements of HetNets. However, as will be evident from the discussion below, most of the efforts have been focused at modeling intra- and inter-tier {\em repulsion} that exists in the BS locations due to cell planning. There is relatively less attention given to modeling user-BS attraction, which is the main focus of this paper. 
 

{1) \em Intra-tier coupling.} One of the conspicuous shortcomings of the PPP model is its inability to model minimum inter-site distance that exists in cellular networks due to cell site planning. This motivated several works in which the BS locations were modeled by {\em repulsive} point processes, such as  \matern hard-core process  \cite{andrews_lte_wifi}, Gauss-Poisson process \cite{Haenggi_gauss_poisson}, Ginibre point process \cite{Haenggi_ginibre}, and determinantal point process \cite{Li_Dhi_DPP}. For completeness, it should be noted that in high shadowing regime, the network topology does {\em appear} Poissonian to the receiver even if it follows a repulsive process~\cite{keeler2014wireless}. This justifies the use of a PPP for modeling BS locations if the propagation channels exhibit sufficiently strong shadowing that is independent across links~\cite{BlaKarC2013,keeler2014wireless}.

{2) \em Inter-tier coupling.} Another conspicuous shortcoming of the $K$-tier HetNet model is the assumption of independence in the locations of the BSs across tiers. While this independence can be justified to some extent between MBSs and user-deployed SBSs (because users do not usually know the MBS topology), it is a bit more questionable for the SBSs deployed by the operators who will tend to concentrate them towards the cell edge away from the MBSs. This has motivated the use of Poisson hole process (PHP) \cite{yazdanshenasan2016poisson} for modeling HetNets \cite{DengHaenggiHeterogeneous2015,yazdanshenasan2016poisson,PP_cbrs}. In this model, the MBSs are first modeled by a PPP. Inhibition zone of a fixed radius is then created around each MBS. The SBS locations are then modeled by a PPP {\em outside} these inhibition zones. This introduces repulsion between the locations of the MBSs and SBSs.  


{3) \em User-SBS coupling.} As discussed already, coupling in the locations of the users and SBSs originate from the deployment of SBSs in the user hotspots. This coupling is at the core of several important user and SBS configurations considered in the 3GPP simulation models for HetNets~\cite{3gppreportr12,3gppreportr13,access2010further}. Some relevant configurations motivated by this coupling are summarized in Table~\ref{table::3gpp_configuration}. Note that while the inter- and intra-tier couplings discussed above were modeled using repulsive point processes, accurate modeling of user-SBS coupling requires the use of point processes that exhibit inter-point attraction. Despite the obvious relevance of this coupling in HetNets, until recently this was almost completely ignored in stochastic geometry-based HetNet models. One exception is \cite{NonUniformDhillon}, which proposed a conditional thinning-based method of biasing the location of the typical user towards the BSs, thus inducing coupling in the BS and user locations. While this provided a good enough first order solution, it lacks generality and is not easily extendible to HetNets. The first work to properly incorporate this user-SBS coupling in a $K$-tier HetNet model is~\cite{DownlinkChiranjib2016,SahaAfshDh2016}, in which the the users were modeled as a PCP (around SBS locations) instead of an independent PPP as was the case in the classical $K$-tier model. There are some other recent works that use PCPs to model SBS and/or user locations. Instead of simply listing them here, we discuss them next in the context of four 3GPP-inspired generative models, which collectively model several key user and SBS configurations of interest in HetNets.%

 \begin{table}
 \centering{
\caption{\small Relevant user and SBS configurations used in 3GPP HetNet models (synthesized from the configurations discussed in \cite[Table~A.2.1.1.2-4]{access2010further}, \cite{3gppreportr12,3gppreportr13}). }\label{table::3gpp_configuration}
\rowcolors{1}{}{lightgray}
\scalebox{0.8}{
\begin{tabular}{c  c c l}
  \hline
 Configuration
 & \vtop{\hbox{\strut User distribution} \hbox{\strut within a macrocell}} & \vtop{\hbox{\strut SBS distribution}\hbox{\strut within a macrocell}} &Comments\\
 1&Uniform&Uncorrelated&Captured by Model $1$\\
 2&Clustered & Correlated, hotspot center &\vtop{\hbox{\strut capacity centric deployment}\hbox{\strut Captured by Model $2$}}\\
 3 &Clustered &Correlated, small cell cluster &\vtop{\hbox{\strut Deployed at user hotspots}\hbox{\strut Cluster size may vary from small to large}\hbox{\strut Captured by Model $3$}} \\
 4 &Uniform &Clustered &\vtop{\hbox{\strut Applies for  pedestrians}\hbox{\strut Captured by Model $4$}}\\
  \hline
\end{tabular}
}}
\end{table}
\subsection{3GPP-inspired generative models using PPP and PCP}
\label{sec::hetnet_model}
As discussed above already, we need to incorporate inter-point interaction in the HetNet models to capture user-SBS coupling accurately. A simple way of achieving that, which is also quite consistent with the 3GPP configurations listed in Table~\ref{table::3gpp_configuration}, is to use PCPs. By combining PCP with a PPP, we can create generative models that are rich enough to model different HetNet configurations of Table~\ref{table::3gpp_configuration}. We discuss these generative models next.
\begin{itemize}
\item \textit{Model 1: SBS PPP, user PPP.} This is the PPP-based $K$-tier {\em baseline} model most commonly used in HetNet literature and is in direct agreement with the 3GPP models with \textit{uniform} user and \textit{uncorrelated} SBS distribution (configuration $1$ in Table~\ref{table::3gpp_configuration}). 
\item \textit{Model 2: SBS PPP, user PCP.} Proposed in our recent work~\cite{DownlinkChiranjib2016,SahaAfshDh2016}, this model can accurately characterize \textit{clustered} users and \textit{uncorrelated} SBSs.  In particular, we model the clustered user and SBS locations jointly  by defining PCP of users around PPP distributed SBSs. This captures the coupling between user and SBS locations. More precisely, this model closely resemblances the 3GPP configuration of single SBS per user hotspot in a HetNet, which is listed as configuration $2$ in Table~\ref{table::3gpp_configuration}.  
\item \textit{Model 3: SBS PCP, user PCP.} The SBS locations exhibit inter-point attraction (and coupling with user locations) when multiple SBSs are deployed in each user hotspot. For modeling such scenarios, two PCPs with the same parent PPP but independent and identically distributed (i.i.d.) offspring point processes can be used to model the user and SBS locations. Coupling is modeled by having the same parent PPP for both the PCPs. We proposed and analyzed this model for HetNets in~\cite{AfshDhiClusterHetNet2016} (models configuration $3$ listed in Table~\ref{table::3gpp_configuration}).
 \item \textit{Model 4: SBS PCP, user PPP.} This scenario can occur in conjunction with the previous one since some of the users may not be a part of the user clusters but are still served by the clustered SBSs. PPP is  a good choice for modeling user locations in this case~\cite{HetPCPGhrayeb2015, MultiChannel2013}. This corresponds to configuration $4$ in Table~\ref{table::3gpp_configuration}.
\end{itemize}   
These generative models are illustrated in \figref{fig::hetnet::model}. Clearly, they collectively encompass a rich set of 3GPP HetNet configurations. In this paper, we unify these four models and develop a general analytical approach for the derivation of downlink coverage probability. Unlike prior works on PCP-based HetNet models that focused exclusively on {\em max-power based association} policy, we will consider {\em max-$\sir$ cell association}, which will require a completely new formalism compared to these existing works. It is worth noting that this work is the first to consider max-$\sir$ based association in PCP enhanced HetNets. More details about the contributions are provided next.


\subsection{Contribution}
\subsubsection{A unified framework with PCP and PPP modeled BSs and users}
Inspired by the user and SBS configurations considered in the 3GPP simulation models for HetNets (summarized in Table \ref{table::3gpp_configuration}), we propose a unified $K$-tier HetNet model in which 
an arbitrary number of BS tiers and a fraction of users is modeled by PCPs. The PCP assumption for the BS tier incorporates spatial coupling among the BS locations. On the other hand, the coupling between user and BS locations is captured when the users are also modeled as a PCP with each cluster having either (1) a BS at its cluster center, or (2) a BS cluster with same cluster center as that of the user cluster. As will be evident soon, the four generative models discussed above (and the four user and SBS configurations listed in Table~\ref{table::3gpp_configuration}) can all be treated as special cases of this general setup. 


\subsubsection{Sum-product functional and coverage probability analysis}


We derive coverage probability (or equivalently $\sir$ distribution) of a typical user for the proposed unified HetNet model under the max-$\sir$ cell association. We demonstrate that the coverage probability for this setup can be expressed as a summation of  a functional over the BS point processes which we define as {\em sum-product functional}. As a part of the analysis, we
characterize this functional for PPP, PCP and its associated offspring point process, thus leading to new results from stochastic geometry perspective that may find broader applications in the field.  After deriving all results in terms of {\em general} PCP, we  specialize them to two cases: when all the clustered BS tiers and users are modeled as (i) \matern cluster process (MCP), and (ii) Thomas cluster process (TCP).

\subsubsection{Limiting behavior}

We also study the limiting behavior of PCP in the context of this model. In particular, we show that when the cluster size tends to infinity: (i) the PCP {\em weakly} converges to a PPP, (ii) the limiting PPP and the parent PPP become independent point processes. Although, to the best of our knowledge, these limiting results have not been reported in the communications literature (due to limited application of PCPs to communication network modeling), it would not be prudent to claim that they are not known/available in some form in the broader stochastic geometry literature. Regardless, as a consequence of this limiting result, we are able to formally demonstrate that the coverage probability obtained under this general framework converges to the well-known closed-form coverage probability result of \cite{dhillon2012modeling} obtained for the baseline PPP-based HetNet model where all the BS tiers and users are modeled as independent PPPs.

One of the key take-aways of this study is the fact that the performance trends in HetNets strongly depend on the network topology and are highly impacted by the spatial coupling between the user and BS locations. While the PPP-based baseline HetNet model provided useful initial design guidelines, it is perhaps time to focus on more realistic models that are in better agreement with the models used in practice, such as the ones in the 3GPP simulation models. Our numerical studies demonstrate several fundamental differences in the coverage probability trends in Models~1-4 when the parameters of the BS and user point processes are changed. 




 \begin{figure*}[t]
  \begin{subfigure}{.24\textwidth}
  \centering
  \includegraphics[width=\linewidth]{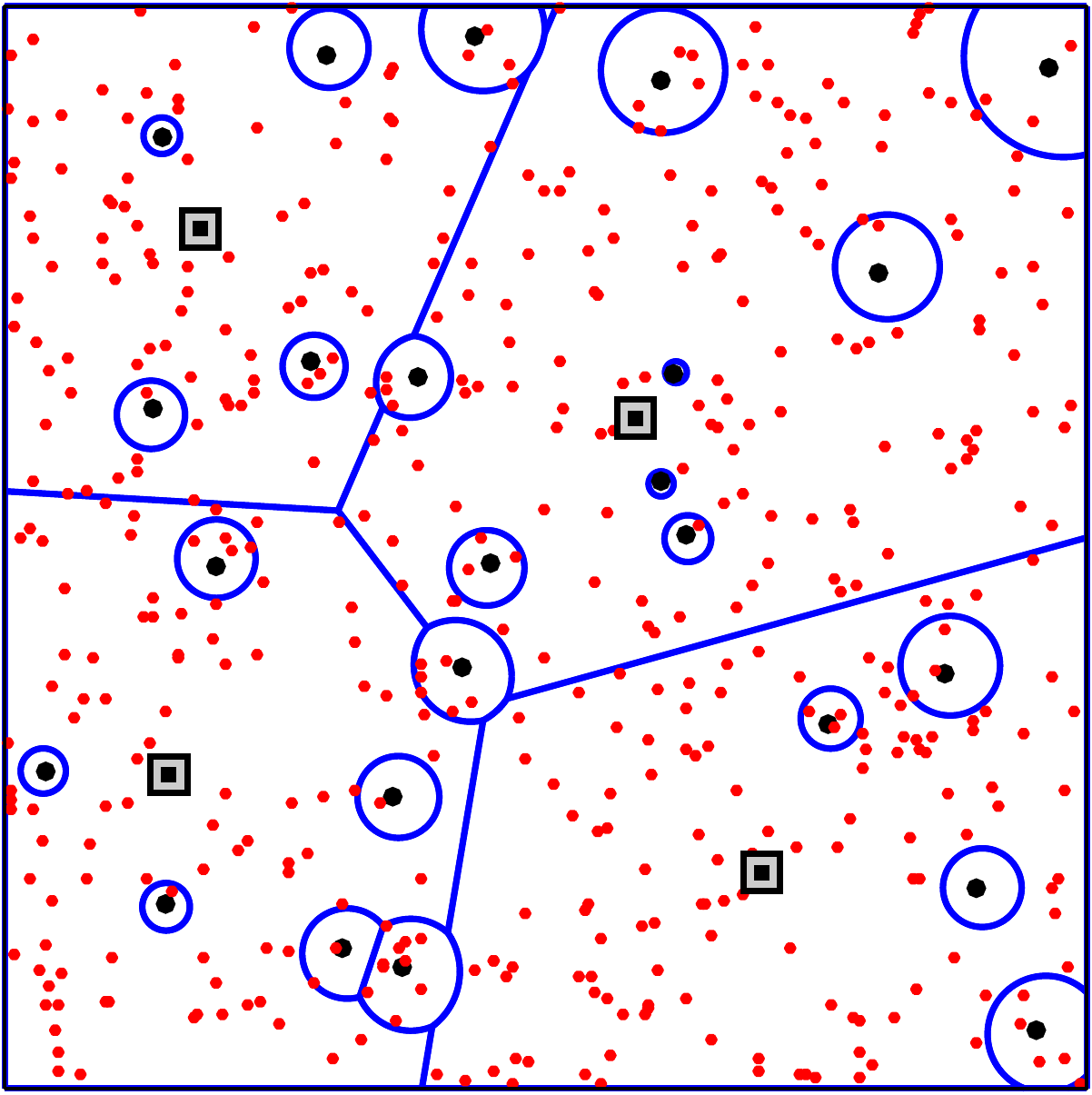}
  \caption{\small Model 1: SBS PPP, user PPP (baseline)}
  \label{fig:sfig1}
\end{subfigure}%
\hspace{0.1cm}
\begin{subfigure}{.24\textwidth}
  \centering
  \includegraphics[width=\linewidth]{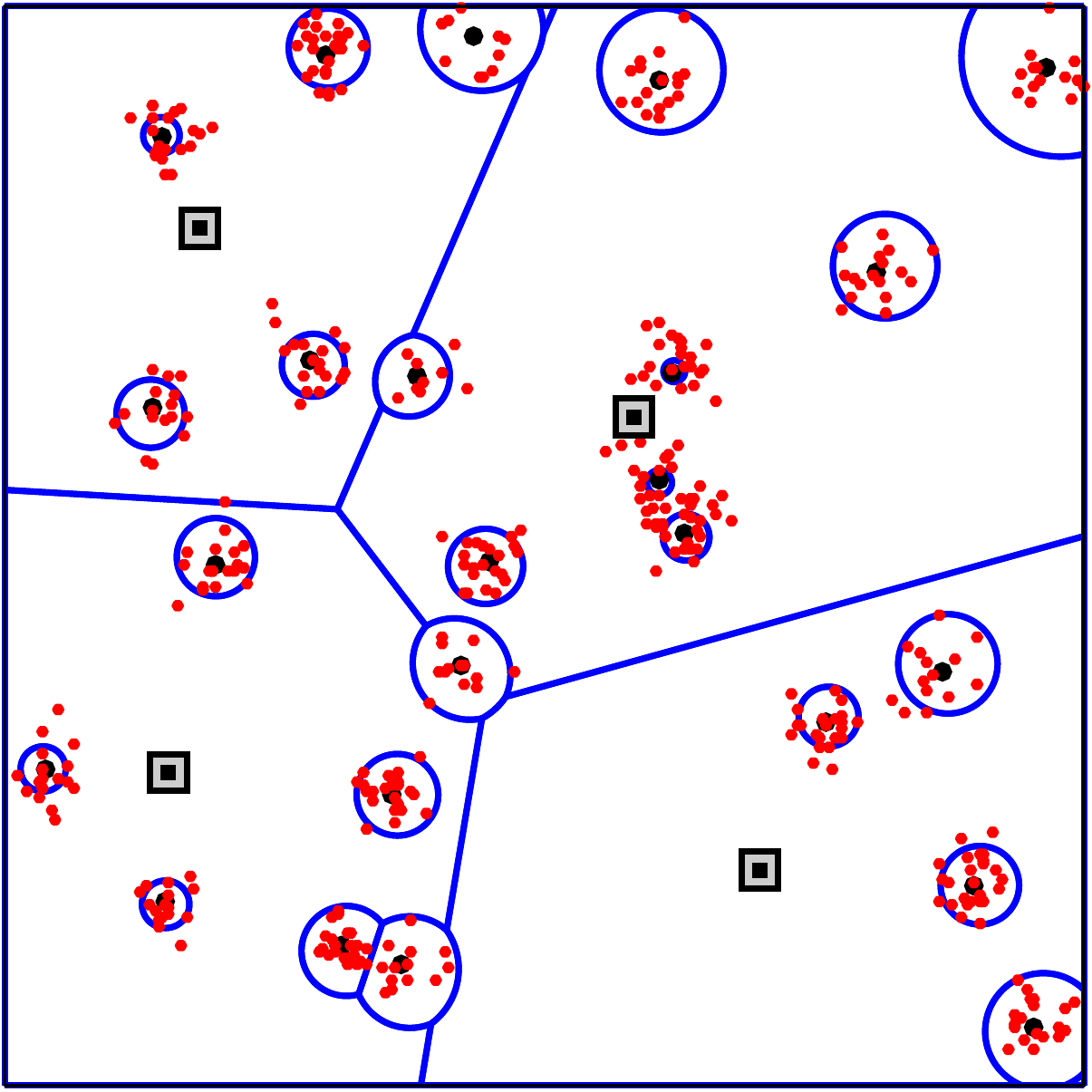}
  \caption{Model 2: SBS PPP, user PCP\ \ \ }
  \label{fig:sfig2}
\end{subfigure}
  \begin{subfigure}{.24\textwidth}
  \centering
  \includegraphics[width=\linewidth]{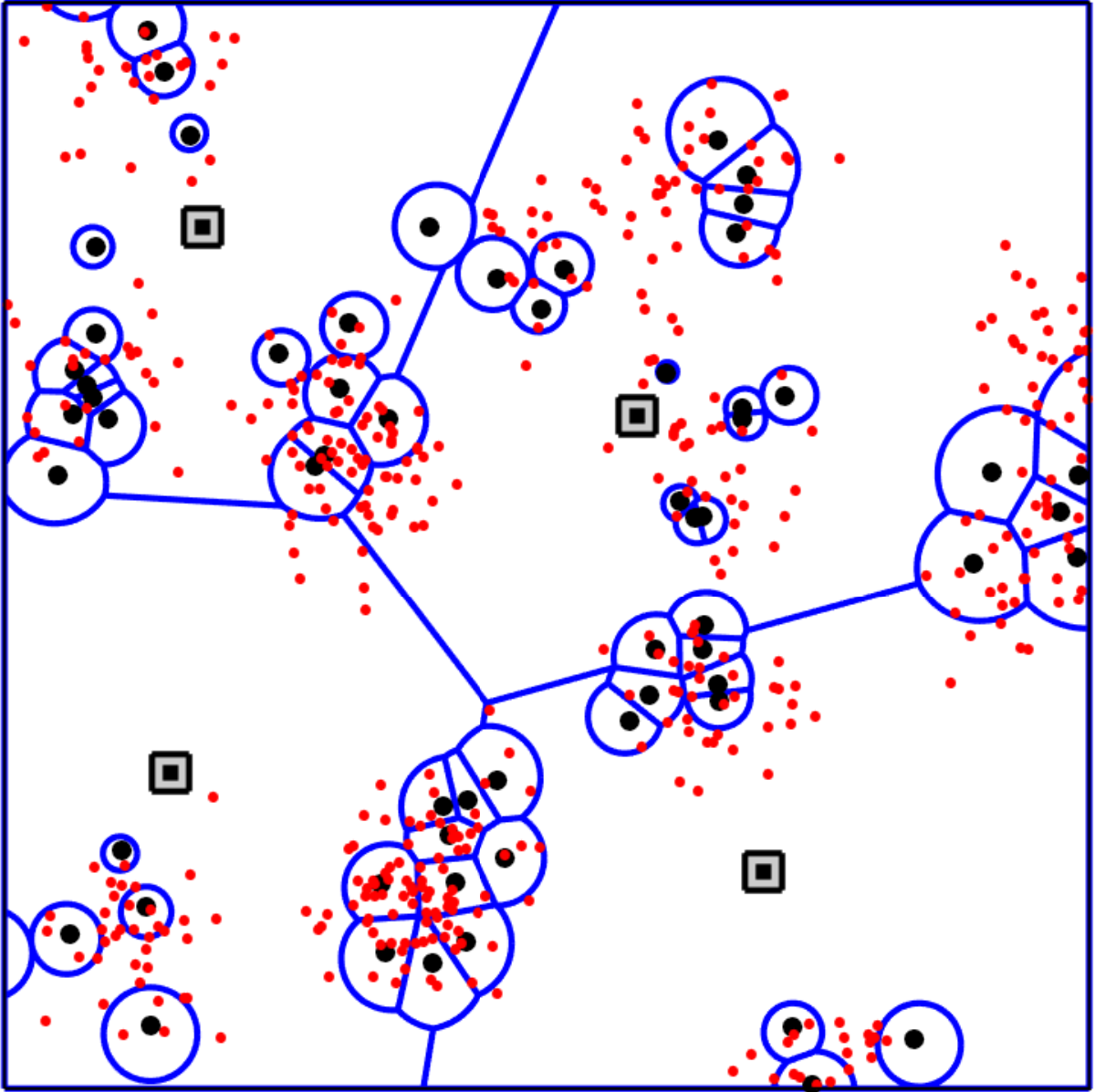}
  \caption{Model 3: SBS PCP, user PCP\ \ \ }
  \label{fig:sfig1}
\end{subfigure}%
\hspace{0.1cm}
\begin{subfigure}{.24\textwidth}
  \centering
  \includegraphics[width=\linewidth]{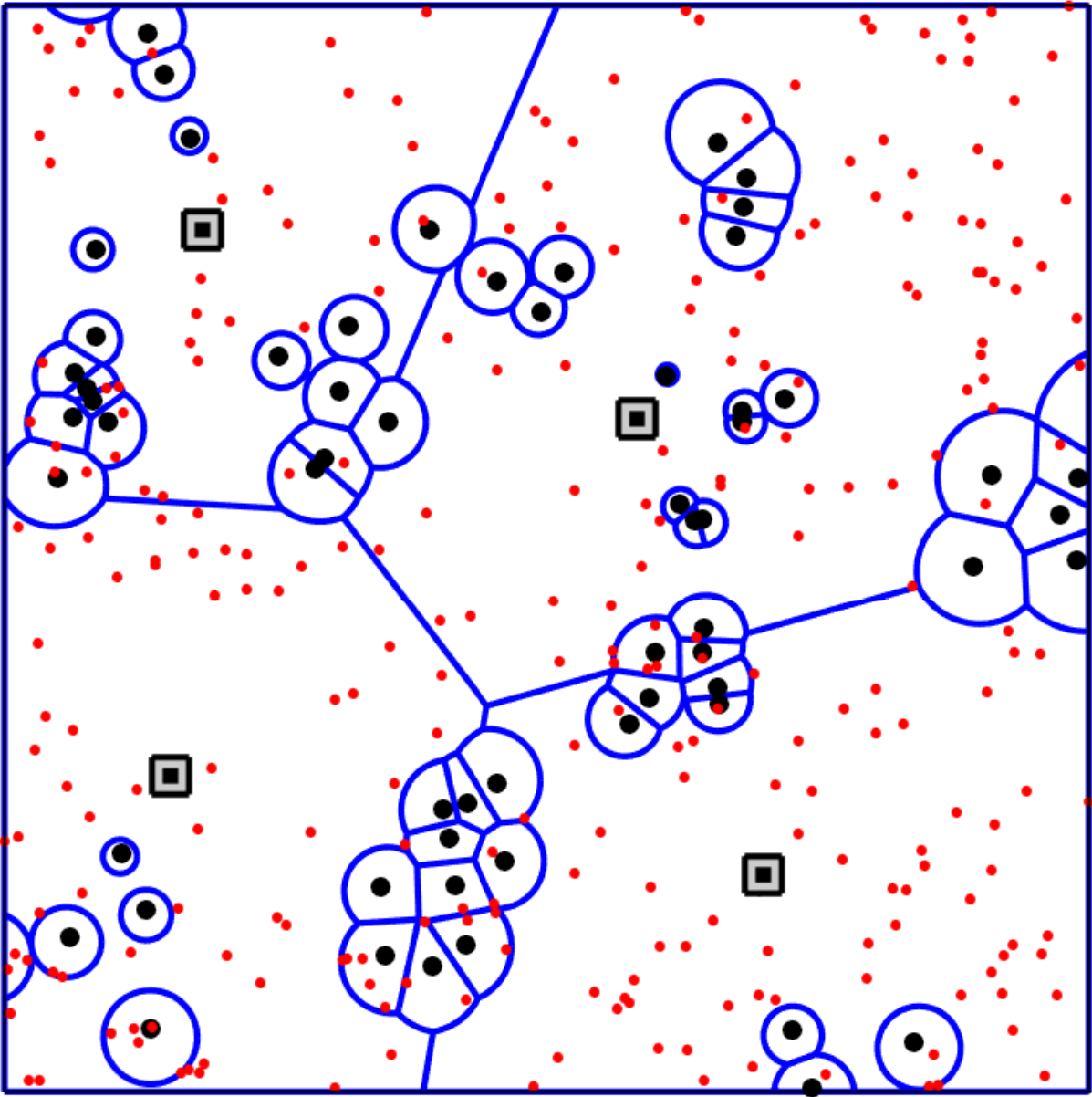}
  \caption{Model 4: SBS PCP, user PPP\ \ \ }
  \label{fig:sfig2}
\end{subfigure}
\caption{\small Illustration of the four generative HetNet models developed by combining PPP and PCP. The black square, black dot and red  dot refer to the MBS, SBS, and users, respectively.}\label{fig::hetnet::model}
\label{fig:fig}
  \end{figure*}

\section{System Model}\label{sec::system_model}
Before we introduce the proposed PCP-based system model for $K$-tier HetNet, we provide a formal introduction to PCP next.
\begin{ndef}[PCP] 
A PCP $\Psi(\lambda_{\rm p}, {f}, p_n)$ can be uniquely defined as:
\begin{equation}\label{eq::pcp_def}
{\Psi= \bigcup_{{\bf z}\in \Phi_{{\rm p}}} {\bf z} + {\cal B}^{\bf z},}
\end{equation} 
where $\Phi_{{\rm p}}$ is the parent PPP of intensity $\lambda_{{\rm p}}$ and $\ncalB^{\bf z}$ denotes the offspring point process corresponding to a cluster center ${\bf z}\in\Phi_{\rm p}$ 
where  $\{{\bf s}\in\ncalB^{\bf z}\}$ is an  i.i.d. sequence of random vectors   with arbitrary probability density function (PDF)  {$f({\bf s})$}. The number of points in $\ncalB^{\bf z}$ is denoted by $N$, where $N\sim p_n$ ($n\in\nbbN$). 
\end{ndef}
 PCP can be viewed as a collection of offspring process ${\ncalB}^{\bf z}$ translated by ${\bf z}$ for each ${\bf z}\in\Phi_{\rm p}$. Then the sequence of points $\{{\bf t}\}\subseteq{\bf z}+\ncalB^{\bf z}$ is conditionally i.i.d. with PDF {$\h({\bf t}|{\bf z}) =f({\bf t}-{\bf z})$}.  A special class of PCP is known as  Neyman-Scott process 
in which $N\sim \mathtt{Poisson}(\bar{m})$.  Throughout this paper,  we will denote the Neyman-Scott process by $\Psi(\lambda_{\rm p},{f},\bar{m})$ and will refer to it as a PCP unless stated otherwise. 
\subsection{$K$-tier HetNet Model}\label{sec::system::model::explained}
We assume a $K$-tier HetNet consisting of $K$ different types of BSs distributed as PPP or PCP. Let $\ncalK_1$ and  $\ncalK_2$ denote the index sets  of the BS  tiers being modeled as PPP and PCP,  respectively, with  $|\ncalK_1\cup\ncalK_2| = K$. 
We denote the point process of the $k^{th}$ BS tier as $\Phi_k$,  where $\Phi_k$ is either a PPP with intensity $\lambda_k$ ($\forall  k\in\ncalK_1$) or a PCP i.e. $\Phi_k(\lambda_{{\rm p}_k},f_k,\bar{m}_k)$ ($\forall k\in\ncalK_2$). 
We assume that each BS of $\Phi_k$ transmits at constant power $P_k$. 
Define $\Phi_{\rm u}$ as the user point process.   Contrary to the common practice in the literature,   $\Phi_{\rm u}$ is not necessarily a PPP independent of the BS locations, rather this scenario will appear as a special case in our analysis. In particular, we consider three different configurations for users: 
\begin{itemize}
\item \caseC~1 ({\em uniform users}): $\Phi_{\rm u}$ is  a  PPP. This corresponds to Models $1$ and $4$ from the previous Section (also see Fig.~\ref{fig:fig}).
\item \caseC~2  ({\em clustered users}): $\Phi_{\rm u}(\lambda_q,f_q,\bar{m}_q)$ is a PCP with parent PPP $\Phi_q$ ($q\in \ncalK_1$), which corresponds to Model $2$ (single SBS deployed in a user hotspot).
\item \caseC~3 ({\em clustered users}): $\Phi_{\rm u}(\lambda_{{\rm p}_q},f_q,\bar{m}_q)$ is a PCP having same parent PPP as that of $\Phi_q$ ($q\in\ncalK_2$), which corresponds to Model $3$ (multiple SBSs deployed at a user hotspot).
\end{itemize}
We perform our analysis for a {\em typical} user which corresponds to a  point selected uniformly at random from $\Phi_{\rm u}$. Since both PPP and PCP are stationary, the typical user is assumed to be located at the origin without loss of generality. 
In \caseS~2 and \caseS~3, the locations of the users and BSs are coupled. Hence, when we select a typical user, we also implicitly select the cluster to which it  belongs. For \caseS~2, let ${\bf z}_0\in\Phi_q\ (q\in\ncalK_1)$ be the location of the BS at the  cluster center of the typical user. For  \caseS~3, let us define the {\em representative} BS cluster $\ncalB^{{\bf z}_0}_{q}\subset \Phi_q\  (q\in\ncalK_2)$ having the  cluster center at ${\bf z}_0$ which is also the cluster center of the typical user located at origin. 
Having defined all three possible configurations/cases of $\Phi_{\rm u}$, we define a set
\begin{align}\label{eq::phi_0_def}
\Phi_0 =\begin{cases} \emptyset;&\text{\caseS~1,}\\
\{{\bf z}_0\};&\text{\caseS~2,}\\
{\bf z}_0+\ncalB^{{\bf z}_0}_q;&{\text{\caseS~3.}}
\end{cases}
\end{align}
This set can be interpreted as the locations of the BSs whose locations are coupled with that of the typical user (alternatively the BSs that lie in the same cluster as the typical user). For the sake of analysis, we remove $\Phi_0$ from $\Phi_q$ and treat it as a separate BS tier (call it the $0^{th}$ tier). Thus, for  \caseS~2,  we remove singleton $\{{\bf z}_0\}$ from $\Phi_q (q\in\ncalK_1)$. 
In  \caseS~3, we remove finite process ${\bf z}_0+\ncalB^{{\bf z}_0}_q$, which is a representative cluster of BSs with properties ($ f_q, \bar{m}_q$) being inherited from $\Phi_q$ ($q\in\ncalK_2$). 
According to Slivnyak's theorem \cite{chiu2013stochastic}, this removal of a point (\caseS~2) or a representative cluster (\caseS~3) does not change the distribution of $\Phi_q$, i.e., $\Phi_q \myeqq{\text{\small d}} \Phi_q\setminus\Phi_0$, where `$\myeqq{\text{\small d}}$' denotes equality in distribution. 
Note that since $\Phi_0$ is constructed from $\Phi_q$ ($q\in\ncalK_1\cup\ncalK_2$), the transmit power of the BS(s) in $\Phi_0$ is $P_0\equiv P_q$. 
Hence, 
the BS point process is a superposition of independent point processes defined as: ${\Phi =   \cup_{k_1\in\ncalK_1}\Phi_{k_1}\cup_{k_2\in\ncalK_2}\Phi_{k_2}\cup\Phi_0,}$
and the corresponding index set is enriched  as: $\ncalK= \ncalK_1\cup\ncalK_2\cup\{0\}$. 
For the ease of exposition, the thermal noise is assumed to be  negligible compared to the  interference power.
Assuming the serving BS is located at ${\bf x}\in\Phi_k$, $\sir({\bf x})$ is defined as:
\begin{align}\label{eq::sir::definition}
\sir({\bf x}) = \frac{P_kh_{\bf x}\|{\bf x}\|^{-\alpha}}{{\ncalI(\Phi_k\setminus\{{\bf x}\})+\sum\limits_{j\in\ncalK\setminus\{k\}}\ncalI(\Phi_j)}},
\end{align} 
where $\ncalI(\Phi_i)  = \sum_{{\bf y}\in\Phi_i}P_i h_{\bf y} \|{\bf y}\|^{-\alpha}$ is the aggregate interference from $\Phi_i$ ($i\in\ncalK$). 
For the channel model, we assume that the signal from a BS at ${\bf y}\in\nbbR^2$ undergoes independent Rayleigh fading, more precisely  $\{h_{\bf y}\}$ is an i.i.d. sequence of random variables, with $h_{\bf y}\sim \exp(1)$, and $\alpha > 2$ is the path-loss exponent. Assuming $\beta_k$ is the $\sir$-threshold defined for $\Phi_k$ for successful connection and the user connects to the BS  that provides maximum $\sir$,  coverage probability is defined as: 
\begin{align}\label{eq::coverage_definition}
\pc &= \nbbP\bigg[\bigcup\limits_{k\in\ncalK}\bigcup\limits_{{\bf x}\in\Phi_k}\{ \sir({\bf x})>\beta_k \}\bigg].
\end{align}
Note that $\beta_0 \equiv \beta_q$ for \caseS~2 and \caseS~3, as discussed above already. The main goal of this paper is to provide exact characterization of $\pc$ for this general model. In the next Section, we derive some intermediate results which will be necessary for this characterization.
\section{Point Process Functionals} 
This is the first main technical section of this paper, where we characterize the  sum-product functional and probability generating functional (PGFL) of a point process $\Psi$ with respect to both its original and reduced Palm distributions, where $\Psi$ can be either a PPP, PCP or its associated offspring process. While PGFLs of point processes are widely-known functionals in stochastic geometry \cite{chiu2013stochastic}, sum-product functionals are not as well-studied. Perhaps the most relevant prior work on sum-product functionals is \cite{SchilcherToumpisHaenggi2016} but it was limited to PPPs. 
These point process functionals will be used in the analysis of coverage probability  under max-$\sir$ connectivity in the next Section. We begin by providing their formal definitions. 
\begin{ndef}[Sum-product functional]\label{def::sum::profuct} Sum-product functional of a point process $\Psi$ is defined in this paper as:
\begin{align}\label{eq::sum::product::defn}
\nbbE\left[\sum\limits_{{\bf x}\in{\Psi}}g({\bf x})\prod\limits_{{\bf y}\in\Psi\setminus\{{\bf x}\}}v({\bf x},{\bf y})\right],
\end{align}
where $g({\bf x}):\nbbR^2\mapsto[0,1]$ and $v({\bf x},{\bf y}):[\nbbR^2\times\nbbR^2]\mapsto[0,1]$ are measurable. 
\end{ndef}
Note that our  definition of the sum-product functional is slightly different from the way it was defined (for PPPs) in \cite{SchilcherToumpisHaenggi2016}. In \eqref{eq::sum::product::defn}, while taking product over $\Psi$, we exclude the point ${\bf x}$ appearing in the outer summation. It will be evident later that this invokes reduced Palm measures of $\Psi$. Also note that the above functional form can be treated as a special case of the functional that appears in the definition of Campbell-Mecke theorem \cite{chiu2013stochastic}. Next we define the  PGFLs of a  point process with respect to its original and   reduced Palm distribution. 
\begin{ndef}[PGFL]
The PGFL of a point process $\Psi$ evaluated at $v({\bf x},{\bf y})$ is defined as:
\begin{align}\label{eq::pgfl::defn}
G(v({\bf x},{\bf y})) = \nbbE\left[\prod\limits_{{\bf y}\in \Psi}v({\bf x},{\bf y})\right],
\end{align} 
where $v({\bf x},{\bf y}):[\nbbR^2\times\nbbR^2]\mapsto [0,1]$ is measurable.  
The PGFL of $\Psi$ under the condition of removing a point of $\Psi$ at ${\bf x}$ or alternatively the PGFL of $\Psi$ under its reduced Palm distribution is defined as:
\begin{align}\label{eq::pgfl::defn}
\widetilde{G}(v({\bf x},{\bf y})) = \nbbE_{\bf x}^!\left[\prod\limits_{{\bf y}\in \Psi}v({\bf x},{\bf y})\right]=\nbbE\left[\prod\limits_{{\bf y}\in \Psi\setminus\{{\bf x}\}}v({\bf x},{\bf y})\right].
\end{align} 
\end{ndef}
Although it is natural to define PGFL of a point process at some $v'({\bf y})$ where $v':\nbbR^2\mapsto[0,1]$ is measurable,  we define PGFL at  $v({\bf x},{\bf y})$, where `${\bf x}$' appears as a dummy variable,  to be consistent with the notation used throughout this paper.
\subsection{Sum-product Functionals}\label{sec::sum::product}
In this Subsection, we characterize the sum-product functionals of different point processes that appear in the expression for coverage probability of a typical user in the next Section. 
  The  sum-product functional  when $\Psi$ is a PPP is presented in the next Lemma.
\begin{lemma}\label{lemm::sumproduct::ppp}The sum-product functional of $\Psi$ when $\Psi$ is a PPP of intensity $\lambda$ is: 
\begin{align}\label{eq::sumproduct::ppp}
\nbbE\left[\sum\limits_{{\bf x}\in{\Psi}}g({\bf x})\prod\limits_{{\bf y}\in\Psi\setminus\{{\bf x}\}}v({\bf x},{\bf y})\right]&=\lambda\int\limits_{\nbbR^2} g({\bf x})\widetilde{G}(v({\bf x},{\bf y})){\rm d}{\bf x},
\end{align}
where $\widetilde{G}(v({\bf x},{\bf y}))$ is the PGFL of $\Psi$ with respect to its reduced Palm distribution and  $\widetilde{G}(v({\bf x},{\bf y}))={G}(v({\bf x},{\bf y})) $. 
\end{lemma}
\begin{IEEEproof}
We can directly apply Campbell-Mecke Theorem \cite{chiu2013stochastic} to evaluate \eqref{eq::sum::product::defn} as:
\begin{align*}
&\nbbE\bigg[\sum\limits_{{\bf x}\in\Psi}g({\bf x})\prod\limits_{{\bf y}\in\Psi\setminus\{{\bf x}\}}v({\bf x},{\bf y})\bigg]
 =  \int\limits_{\nbbR^2}g({\bf x})\nbbE_{\bf x}^! \prod\limits_{{\bf y}\in\Psi}v({\bf x},{\bf y})\Lambda({\rm d}{\bf x}) = \int\limits_{\nbbR^2} g({\bf x})\widetilde{G}(v({\bf x},{\bf y}))\Lambda({\rm d}{\bf x}),
\end{align*}
where $\Lambda(\cdot)$ is  the intensity measure of $\Psi$ and  $\widetilde{G}(\cdot)$ denotes the PGFL of $\Psi$ under its reduced Palm distribution. When $\Psi$ is  homogeneous PPP, $\Lambda({\rm d}{\bf x})=\lambda\:{\rm d}{\bf x}$ and $\widetilde{G}(v({\bf x},{\bf y})) =G(v({\bf x},{\bf y})) = \nbbE\prod\limits_{{\bf y}\in\Psi} v({\bf x},{\bf y})$, by Slivnyak's theorem~\cite{chiu2013stochastic}.
\end{IEEEproof}

Sum-product functional of $\Psi$ when $\Psi$ is a PCP requires  more careful treatment since selecting a point from ${\bf x}\in\Psi$ implies selecting a tuple $({\bf x},{\bf z})$, where ${\bf z}$ is the cluster center of ${\bf x}$. Alternatively, we can assign a two-dimensional mark ${\bf z}$ to each point ${\bf x}\in\Psi$ such that ${\bf z}$ is the cluster center of ${\bf x}$. Then $({\bf x},{\bf z})$ is a point from the marked point process $\hat{\Psi}\subset\nbbR^2\times\nbbR^2$. It should be noted that $ \hat{\Psi}$ is
simply an alternate representation of  $\Psi$, which will be useful
in some proofs in  this Section. Taking $A,B\subset \nbbR^2$, its intensity measure can be expressed as: $\Lambda(A, B) = $
\begin{align*}
\nbbE\bigg[\sum\limits_{\substack{\small ({\bf x},{\bf z})\in \hat{\Psi}}}{\bf 1}\big({\bf x}\in A, {\bf z}\in B\big)\bigg]
\myeq{a}\nbbE\bigg[\sum\limits_{{\bf z} \in\Phi_{{\rm p}}\cap B}\bar{m}\int\limits_{{\bf x}\in A}\h({\bf x}|{\bf z}){\rm d}{\bf x}\bigg]=\bar{m}\lambda_{{\rm p}}\iint\limits_{\substack{{\bf z}\in B,{\bf x}\in A}}\h({\bf x}|{\bf z}){\rm d}{\bf x}\:{\rm d}{\bf z},
\end{align*}
where in step (a), the expression under summation is the  intensity of ${\bf z}+\ncalB^{{\bf z}}$, i.e., the offspring process with cluster center at ${\bf z}$. The last step follows from   Campbell's theorem \cite{chiu2013stochastic}. 
Hence, \begin{equation}\label{eq::Lambda_pcp}
\Lambda({\rm d}{\bf x},{\rm d}{\bf z})= \lambda_{{\rm p}}\bar{m}\h({\bf x}|{\bf z})\:{\rm d}{\bf z}\:{\rm d}{\bf x}.  
\end{equation} 
We now evaluate the sum-product functional of PCP in the next Lemma. 
\begin{lemma}\label{lemm::sumproduct::pcp}The sum-product functional of $\Psi$ when $\Psi$ is a PCP  can be expressed as follows: 
\begin{align}\label{eq::sumproduct::pcp}
\nbbE\left[\sum\limits_{{\bf x}\in{\Psi}}g({\bf x})\prod\limits_{{\bf y}\in\Psi\setminus\{{\bf x}\}}v({\bf x},{\bf y})\right]=\iint\limits_{\nbbR^2\times\nbbR^2} g({\bf x})\widetilde{G}(v({\bf x},{\bf y})|{\bf z})\Lambda({\rm d}{\bf x},{\rm d}{\bf z}),
\end{align}
where
\begin{equation}\label{eq::reduced::palm::pcp}
\widetilde{G}(v({\bf x},{\bf y})|{\bf z}) = {G}(v({\bf x},{\bf y}))\widetilde{G}_c(v({\bf x},{\bf y})|{\bf z})
\end{equation}
 denotes the PGFL of $\Psi$ when a  point ${\bf x}\in\Psi$ with cluster center at ${\bf z}$ is removed from $\Psi$. $G(\cdot)$  is the PGFL of $\Psi$
  and $\widetilde{G}_c(\cdot|{\bf z})$ is the PGFL of ${\bf z}+\ncalB^{\bf z}$, which is a cluster of $\Psi$  centered at ${\bf z}$ under its reduced Palm distribution. 
\end{lemma}
\begin{IEEEproof} 
 Starting from \eqref{eq::sum::product::defn}, we apply Campbell-Mecke theorem on $\hat{\Psi}$ as follows:
\begin{align*}
 \nbbE\bigg[\sum\limits_{({\bf x},{\bf z})\in\hat{\Psi}}g({\bf x})\prod\limits_{({\bf y},{\bf z}')\in\hat{\Psi}\setminus({\bf x},{\bf z})}v({\bf x},{\bf y})\bigg]=
\iint\limits_{\nbbR^2\times \nbbR^2} \nbbE_{({\bf x},{\bf z})}^!\bigg[g({\bf x})\prod\limits_{({\bf y},{\bf z}')\in\hat{\Psi}}v({\bf x},{\bf y})\bigg]\Lambda({\rm d}{\bf x},{\rm d}{\bf z}).
\end{align*}
The Palm expectation in the last step can be simplified as:
\begin{align*}
&\nbbE_{({\bf x},{\bf z})}^!\bigg[g({\bf x})\prod\limits_{({\bf y},{\bf z}')\in\hat{\Psi}}v({\bf x},{\bf y})\bigg]
=g({\bf x})\nbbE\bigg[\prod\limits_{{\bf y}\in\Psi\setminus({\bf z}+\ncalB^{{\bf z}}_k)}v({\bf x},{\bf y})\prod\limits_{{\bf y}\in({\bf z}+\ncalB^{{\bf z}}_k)\setminus\{{\bf x}\}}v({\bf x},{\bf y})\bigg]\\
&\myeq{a}g({\bf x})\nbbE\bigg[\prod\limits_{{\bf y}\in\Psi\setminus({\bf z}+{\ncalB}^{\bf z})}v({\bf x},{\bf y})\bigg]\nbbE\bigg[\prod\limits_{{\bf y}\in({\bf z}+\ncalB^{{\bf z}})\setminus\{{\bf x}\}}v({\bf x},{\bf y})\bigg]\myeq{b}g({\bf x})\nbbE\bigg[\prod\limits_{{\bf y}\in\Psi}v({\bf x},{\bf y})\bigg]\nbbE^{!}_{\bf x}\bigg[\prod\limits_{{\bf y}\in({\bf z}+\ncalB^{{\bf z}})}v({\bf x},{\bf y})\bigg],
\end{align*}
where (a) follows from the independence of the  processes ${\bf z}+\ncalB^{\bf z}$ and ${\Psi}\setminus({\bf z}+\ncalB^{\bf z})$
 and (b) follows from  Slivnyak's theorem for PCP, i.e. $\Psi\myeqq{\text{\small d}}\Psi\setminus({\bf z}+{\ncalB}^{\bf z})$~\cite{ganti2009interference}. Substituting the PGFLs as $\nbbE\prod\limits_{{\bf y}\in\Psi} v({\bf x},{\bf y})=G(v({\bf x},{\bf y}))$, and $\nbbE_{\bf x}^!\prod\limits_{{\bf y}\in{\bf z}+\ncalB^{\bf z}}v({\bf x},{\bf y})=\widetilde{G}_c(v({\bf x},{\bf y})|{\bf z}) $,
we get the final result.
\end{IEEEproof}

The similar steps for the evaluation of the sum-product functional can not be followed when $\Psi$ is a finite point process,
specifically, $\Psi={\bf z}+\ncalB^{\bf z}$, the cluster of a randomly chosen
point ${\bf x} \in \Psi$ centered at ${\bf z}$. 


\begin{lemma}\label{lemm::sumproduct::finite}The sum-product functional of $\Psi$ when $\Psi={\bf z}+\ncalB^{\bf z}$, i.e.,  the offspring point process of a PCP centered at ${\bf z}$
can be expressed as follows: 
\begin{multline}\label{eq::sumproduct::finite}
\nbbE\left[\sum\limits_{{\bf x}\in{\Psi}}g({\bf x})\prod\limits_{{\bf y}\in\Psi\setminus\{{\bf x}\}}v({\bf x},{\bf y})\right]= \int\limits_{\nbbR^2}g({\bf x})\exp \Big(- \bar{m} \int_{\R^2}(1-v({\bf x},{\bf y})) \h({\bf y}|{\bf z}) {\rm d} {\bf y}\Big)\\
\times \Big(\bar{m}   \int_{\R^2} v({\bf x},{\bf y}) \h({\bf y}|{\bf z}) {\rm d} {\bf y} +1  \Big) \h({\bf x}|{\bf z}){\rm d}{\bf x}.
%
%
\end{multline}
\end{lemma}
\begin{IEEEproof}
Note that $\Psi$ is conditioned to have at least one point (the one located at ${\bf x}$) and the number of points in $\Psi$ follows a weighted distribution, $\widetilde{N} \sim \frac{n p_n}{\bar{m}}$ ($n\in\nbbZ^+$) \cite{chiu2013stochastic}.  Now, starting from \eqref{eq::sum::product::defn},
\begin{align*}
&\int\limits_{\ncalN}\sum\limits_{{\bf x\in\psi}}g({\bf x})\prod\limits_{{\bf y}\in\psi\setminus\{{\bf x}\}}v({\bf x},{\bf y})P({\rm d }\psi)
\myeq{a}{\sum\limits_{n=1}^{\infty}\int\limits_{{\ncalN}_n}\sum\limits_{{\bf x}\in\psi}g({\bf x})\prod\limits_{\substack{{\bf y}\in\psi\setminus\{{\bf x}\}}} v({\bf x},{\bf y})P({\rm d}{\psi})}
\\&=\sum\limits_{n=1}^{\infty}\ \idotsint\limits_{[{\bf x}_1,\dots,{{\bf x}_n}]\in\nbbR^{2n}}\sum\limits_{i=1}^ng({\bf x}_i)\bigg[\prod\limits_{\substack{j=1,\\j\neq i}}^{n}v({\bf x}_i,{\bf x}_j)\h({\bf x}_j|{\bf z}){\rm d}{\bf x}_j\bigg]  \h({\bf x}_i|{\bf z}){\rm d}{\bf x}_i \frac{n p_n}{\bar{m}}
\\&=\sum\limits_{n=1}^{\infty}n\int\limits_{\nbbR^2}g({\bf x}) \left(\ \int\limits_{\nbbR^2}v({\bf x},{\bf y})\h({\bf y}|{\bf z}){\rm d}{\bf y}\right)^{n-1} \h({\bf x}|{\bf z}){\rm d}{\bf x}\: n\frac{p_n}{\bar{m}},
\end{align*}
where  $\ncalN$ denotes the space of locally finite and simple point sequences in $\nbbR^2$. In (a), $\ncalN$ is partitioned into $\{\ncalN_{n}: n\geq 1 \}$ where $\ncalN_n$ is the collection of point sequences having $n$ points and  $\psi$ denotes a realization of $({\bf z}+\ncalB^{\bf z})$. 
 Under the condition of removing a point $\bf x$ from $({\bf z}+\ncalB^{\bf z})$, this point process  will have at least one point. Hence, the number of points in $({\bf z}+\ncalB^{\bf z})$ will follow the weighted distribution: ${\tilde{N}}\sim \frac{np_{n}}{\bar{m}}$ ($n\in\nbbZ^+$). {The final expression of $\widetilde{G}_c$ can be obtained by substituting $p_n (\forall\ n\in\nbbN)$ by the probability mass function (PMF) of Poisson distribution followed by basic algebraic manipulations.}
\end{IEEEproof}
\subsection{Probability Generating Functional}\label{sec::pgfl::def}
In this Section, we evaluate the PGFLs of different point processes that appeared in the expressions of the sum-product functionals in the previous Section. While the PGFLs of the PPP and PCP are known~\cite{haenggi2012stochastic}, we list them in the next Lemma for completeness.  
\begin{lemma}\label{lemm::pgfl::ppp::pcp}
 The PGFL of $\Psi$ when $\Psi$ is a PPP of intensity $\lambda$ is given by:
\begin{equation}\label{eq::pgfl::ppp}
G(v({\bf x},{\bf y})) = \exp\left(-\lambda\int_{\nbbR^2}(1-v({\bf x},{\bf y})){\rm d}{\bf y}\right).
\end{equation}
When $\Psi$ is a PCP, the PGFL of  $\Psi\  (\lambda_{\rm p},f,\bar{m})$  is given by:
\begin{align}\label{eq::pgfl::pcp}
{{G}(v({\bf x},{\bf y}))=\exp\left(-\lambda_{\rm p}\int\limits_{\nbbR^2}\left(1-\exp\left(-\bar{m}\bigg(1-\int\limits_{\nbbR^2}v({\bf x},{\bf y})\h({\bf y}|{\bf z}){\rm d}{\bf y}\bigg)\right)\right){\rm d}{\bf z}\right)}.
\end{align} 
\end{lemma}
\begin{IEEEproof} Please refer to \cite[Theorem 4.9, Corollary 4.13]{haenggi2012stochastic}.
\end{IEEEproof}
We have pointed out in Lemma~\ref{lemm::sumproduct::ppp} that the PGFLs with respect to the original and reduced Palm distributions are the same when $\Psi$ is a PPP. However, this is not true for PCP. It was shown in Lemma~\ref{lemm::sumproduct::pcp} that when $\Psi$ is a PCP, the PGFL of $\Psi$ ($\lambda_{\rm p},f,\bar{m}$) with respect to its reduced Palm distribution is given by the product of its PGFL $G(v({\bf x},{\bf y}))$ and $\widetilde{G}_c(v({\bf x},{\bf y})|{\bf z})$, where $\widetilde{G}_c(v({\bf x},{\bf y})|{\bf z})$ is the PGFL of ${\bf z}+\ncalB^{\bf z}$ with respect to its reduced Palm distribution. We characterize   ${G}_c(v({\bf x},{\bf y})|{\bf z})$ and $ \widetilde{G}_c(v({\bf x},{\bf y})|{\bf z})$  in the next Lemma. 
\begin{lemma}\label{lemm::pgfl::cluster}
 The PGFL  of $\Psi$ when $\Psi = {\bf z}+\ncalB^{\bf z}$ conditioned on the removal of a point at ${\bf x}$ is:
 \begin{align}\label{eq::pgfl_cluster_reduced}
{\widetilde{G}_c(v({\bf x},{\bf y})|{\bf z})  ={G}_c(v({\bf x},{\bf y})|{\bf z}) ,}
\end{align} 
where $G_c(v({\bf x},{\bf y}))$ is the PGFL of ${\bf z}+\ncalB^{\bf z}$ which is given by:
\begin{align}\label{eq::pgfl::reduced::cluster}
{{G}_c(v({\bf x},{\bf y})|{\bf z})= \exp\bigg(-\bar{m}\bigg(1-\ \int\limits_{\nbbR^2}v({\bf x},{\bf y}) \h({\bf y}|{\bf z}){\rm d}{\bf y}\bigg)\bigg).}
\end{align}
\end{lemma}
\begin{IEEEproof}The PGFL of $\Psi$ with respect to its reduced Palm distribution can be expressed as:
\begin{align*}
&\widetilde{G}_c(v({\bf x},{\bf y})|{\bf z})  = \int\limits_{\ncalN}\prod\limits_{{\bf y}\in\psi}v({\bf x},{\bf y})P^!_{\bf x}({\rm d }\psi)
  \myeq{a}\sum\limits_{n=1}^{\infty}\int\limits_{{\ncalN}_n}\prod\limits_{{\bf y}\in\psi\setminus\{{\bf x}\}}^n v({\bf x},{\bf y})P({\rm d}{\psi})
\\&=\sum\limits_{n=1}^{\infty}\ \idotsint\limits_{[{\bf y}_1,\dots,{{\bf y}_{n-1}}]\in\nbbR^{2n-2}}\prod\limits_{\substack{i=1}}^{n-1}v({\bf x},{\bf y}_i)\h({\bf y}_i|{\bf z}){\rm d}{\bf y}_i \frac{np_{n}}{\bar{m}}
=\sum\limits_{n=1}^{\infty}\left(\ \int\limits_{\nbbR^2}v({\bf x},{\bf y})\h({\bf y}|{\bf z}){\rm d}{\bf y}\right)^{n-1}\: n\frac{p_{n}}{\bar{m}},
\end{align*}
where (a) follows on similar lines of step (a) in the proof of Lemma~\ref{lemm::sumproduct::finite}.  This means we have partitioned $\ncalN$ in the same way as we did in the proof of Lemma~\ref{lemm::sumproduct::finite}. Since we condition on a point ${\bf x}$ of $\Psi$ to be removed, it implies that $\Psi$ will have at least one point. Hence, the number of points in $\Psi$ will follow the weighted distribution: ${\tilde{N}}\sim \frac{n p_n}{\bar{m}}$ (as was the case in Lemma~\ref{lemm::sumproduct::finite}). Similarly, the PGFL of $\Psi = {\bf z}+\ncalB^{\bf z}$ with respect to its original distribution can be obtained by
\begin{align}
{G}_c(v({\bf x},{\bf y})|{\bf z})=\sum\limits_{n=0}^{\infty}\left(\ \int\limits_{\nbbR^2}v({\bf x},{\bf y})\h({\bf y}|{\bf z}){\rm d}{\bf y}\right)^{n}\: p_n.
\end{align}
Substituting $p_n (\forall\ n\in\nbbN)$ by the {PMF} of Poisson distribution, we get the desired expression. 
\end{IEEEproof}

\begin{remark}\label{rem::pgfl::offspring::pp}We observe that the PGFLs of the offspring point process associated with the PCP are the same under the original and the reduced Palm distribution. From the proof of Lemma~\ref{lemm::pgfl::cluster}, it is evident that this result is a consequence of the fact that the number of points in the offspring point process is Poisson~\cite[Section 5.3]{chiu2013stochastic}. 
\end{remark}

\section{Coverage Probability Analysis}
\label{sec::coverage::probability}
This is the second main technical section of this paper, where we evaluate the coverage probability of a typical user in the unified HetNet model which was defined in \eqref{eq::coverage_definition}. Using the results for the point process functionals derived in the previous Section, we first characterize the coverage probability when clustered nodes (users and/or BSs) are modeled as  Neyman-Scott cluster process, and then specialize  our result to the case when clustered users and/or BSs are distributed according to MCPs and TCPs.

\subsection{Neyman-Scott cluster process}
We now provide our main result of downlink  coverage probability of a typical user for the general $K$-tier HetNet setup defined in Section~\ref{sec::system::model::explained} in the following Theorem. 

\begin{theorem}\label{thm::coverage} Assuming that the typical user connects to the BS providing maximum $\sir$ and $\beta_k>1,\ \forall\ k\in\ncalK$, coverage probability can be expressed as follows:
\begin{align}\label{eq::coverage_main_result}
\pc &=\sum\limits_{k\in\ncalK}\pc_{k}=\sum\limits_{k\in\ncalK} \nbbE\bigg[\sum\limits_{{\bf x}\in\Phi_k} \prod\limits_{j\in\ncalK\setminus\{k\}}G_j(v_{k,j}({\bf x},{\bf y}))\prod\limits_{{\bf y}\in \Phi_k\setminus
\{{\bf x}\}}v_{k,k}({\bf x},{\bf y})\bigg]
\end{align}
with
\begin{align}\label{eq::v_function}
v_{i,j}({\bf x},{\bf y}) = \frac{1}{1+\beta_i\frac{P_j}{P_i}\big(\frac{\|{\bf x}\|}{\|{\bf y}\|}\big)^{\alpha}},
\end{align}
where $\pc_{k}$ denotes per-tier coverage probability, more precisely, the joint probability of the event that the serving BS belongs to $\Phi_k$ and the typical user is under coverage, and  $  G_j(\cdot), \forall j\in {\cal K}_1\cup {\cal K}_2$ is given by~Lemma~\ref{lemm::pgfl::ppp::pcp}. 
\end{theorem}
\begin{IEEEproof}
See Appendix~\ref{app::thm::coverage}.
\end{IEEEproof}
%

\begin{remark}[Coverage probability is the summation of $K+1$ sum-product functionals] In \eqref{eq::coverage_main_result},  $\pc$ is the summation of $(K+1)$  {per-tier coverage probabilities}, due to the contribution of $(K+1)$ tiers in $\Phi= \bigcup\limits_{k\in\ncalK}\Phi_{k}$.  Recalling  Definition~\ref{def::sum::profuct},  $\pc_k$ is  in the form of sum-product functional over $\Phi_k$, with $g({\bf x})\equiv \prod_{j\in\ncalK\setminus\{k\}} G_j(v_{k,j}({\bf x},{\bf y}))$ and $v({\bf x},{\bf y})\equiv v_{k,k}({\bf x},{\bf y})$ in \eqref{eq::sum::product::defn}.
\end{remark}

In the previous Section, we have computed the sum-product functional over PPP, PCP and the offspring point process in terms of  arbitrary measurable functions  $g(\bf x)$ and $v({\bf x},{\bf y})$. We directly apply these results to compute $\pc_k$. 
We first provide the expression of PGFL of $\Phi_0$ evaluated at $v_{k,0}({\bf x},{\bf y})$. Depending on the construction of $\Phi_0$ based on three different configurations of $\Phi_{\rm u}$ (refer to \eqref{eq::phi_0_def}), we will have different expressions of $G_0(\cdot)$. 
\begin{lemma}\label{lemm::pgfl_phi_0} The PGFL of  $\Phi_0$ is given by:
\begin{itemize}
\item \caseS~1:\ ${G}_0(v_{k,0}({\bf x},{\bf y}))=1,$
\item \caseS~2:\ $ {G}_0(v_{k,0}({\bf x},{\bf y}))=  \int_{\nbbR^2}\frac{1}{1+   \frac{ P_0 \beta_k}{P_k} \|{\bf x}\|^{\alpha} \|{\bf y}\|^{-\alpha} }  f_0({\bf y}) {\rm d}{\bf  y},$
\item \caseS~3: ${G}_0(v_{k,0}({\bf x},{\bf y}))= \int_{\nbbR^2} {G}_{c_0}(v_{k,0}({\bf x},{\bf y})|{\bf z}')  f_0({\bf z}') {\rm d} {\bf z}',$
\end{itemize}
where $G_c({\cdot|{\bf z}})$ is given by Lemma~\ref{lemm::pgfl::cluster}.
\end{lemma}
\begin{IEEEproof}In \caseS~1,  $\Phi_0$ is a null set if users are distributed according to a PPP, and hence $G_0(v_{k,0}({\bf x},{\bf y}))=1$. In \caseS~2, where users are distributed as a PCP with parent PPP $\Phi_j$ $(j\in \ncalK_1)$,  
\begin{align}
G_0(v_{k,0}({\bf x},{\bf y}))=\int\limits_{\nbbR^2}v_{k,0}({\bf x},{\bf y})f_0({\bf y}){\rm d}{\bf y}.
\end{align}
{
 In \caseS~3,   $\Phi_{0}=\ncalB_j^{{\bf z}_0}$ is a  cluster of $\Phi_j$ ($j\in\ncalK_2$) centered at ${\bf z}_0$}.   Its  PGFL is provided by Lemma~\ref{lemm::pgfl::cluster}, and the final result is obtained by taking expectation  over ${\bf z}_0\sim f_0({\bf z}_0)$.
\end{IEEEproof}
Having characterized the PGFLs of $\Phi_k$ $\forall\ k\in\ncalK$, we evaluate $\pc_k$ in the following Lemmas. 


\begin{lemma}\label{lem:per-tier-cov-pcp}  When the BS tier $\Phi_k$ is a PCP, i.e., $k\in\ncalK_2$, per-tier coverage can be expressed as:
\begin{align}\label{eq::per::tier::cov::pcp}
\pc_k = \iint\limits_{\nbbR^2\times\nbbR^2} {G}_{k}(v_{k,k}({\bf x},{\bf y})) \widetilde{G}_{c_k}(v_{k,k}({\bf x},{\bf y})|{\bf z})\prod_{j\in\ncalK\setminus\{k\}}{G}_j(v_{k,j}({\bf x},{\bf y}))\Lambda_k({\rm d}{\bf x}, {\rm d} {\bf z}),\ k\in\ncalK_2,
\end{align}where 
$\Lambda_k({\bf x},{\bf z})$ is given by \eqref{eq::Lambda_pcp}, 
   $\widetilde{G}_{c_k}(\cdot)$ is obtained by Lemmas~\ref{lemm::pgfl::cluster}. $G_j(\cdot)$ and $G_k(\cdot)$ are 
given by Lemma~\ref{lemm::pgfl::ppp::pcp}.
\end{lemma}
\begin{IEEEproof}
The result is obtained by the direct application of Lemma~\ref{lemm::sumproduct::pcp}.
\end{IEEEproof}
%
%
%
%
%
\begin{remark}When $\Phi_j$ is a PPP, i.e., $j\in\ncalK_1$, $G_{j}(v_{k,j}({\bf x},{\bf y}))$ presented in Lemma~\ref{lemm::pgfl::ppp::pcp}  can be further simplified as:
\begin{align}\label{eq::pgfl_ppp_applied}
{G}_j(v_{k,j}({\bf x},{\bf y}))=\exp\bigg(-\pi \lambda_j \left(\frac{P_j \beta_k}{P_k}\right)^{\frac{2}{\alpha}} \|{\bf x}\|^2  C(\alpha)\bigg); \quad \forall j\in {\cal K}_1,
\end{align}
with $C(\alpha) = \frac{\alpha}{2 \pi} {\sin(\frac{2 \pi} {\alpha})}$. See \cite[Theorem 1]{dhillon2012modeling} for an elaborate proof. 
\end{remark}
In the next Lemma, we present per-tier coverage probability $\pc_k$ $(k\in {\cal K}_1)$.

\begin{lemma}\label{lemm::per_tier_coverage_ppp}  When the BS tier is a PPP, per-tier coverage can be expressed as:
\begin{align}\label{eq::per::tier::cov::ppp}
\pc_k = \lambda_k\int\limits_{\nbbR^2} \prod_{j\in\ncalK}{G}_j(v_{k,j}({\bf  x},{\bf  y})){\rm d}{\bf x}, \qquad k\in\ncalK_1, 
\end{align}
where $G_j(\cdot)$ is obtained by \eqref{eq::pgfl_ppp_applied} for $j\in\ncalK_1$. When $j\in\ncalK_2$, $G_j(\cdot)$ is given by Lemma~\ref{lemm::pgfl::ppp::pcp}. 
\end{lemma}
\begin{IEEEproof}
The result is obtained by the direct application of Lemma~\ref{lemm::sumproduct::ppp}. 
\end{IEEEproof}
Having characterized per-tier coverage $\pc_{k}$ for $k\in\ncalK_1\cup\ncalK_2$, we are left with the evaluation of $\pc_0$ which we do next. Similar to Lemma~\ref{lemm::pgfl_phi_0}, we will have three different cases for $\pc_0$ owing to  different user configurations. 
\begin{lemma}\label{lemm::tier0}
{$\pc_0$ can be expressed as follows.}
\begin{align}
{\tt P}_{{\rm c}_0}=\begin{cases}0 & \text{when $\Phi_0=\emptyset$ (\caseS~1)}\\
\int_{\nbbR^2} \prod\limits_{j\in\ncalK\setminus\{0\}}G_j(v_{0,j}({\bf z}_0,{\bf y})) f_0({\bf z}_0) {\rm d} {\bf z}_0 &\text{when $\Phi_0=\{{\bf z}_0\}$ (\caseS~2),}\\
\int_{\nbbR^2}  \int_{\nbbR^2} \exp \bigg(-\bar{m}_0 \bigg( \int_{\nbbR^2}    \Big( 1-   v_{0,0}({\bf x},{\bf y}) \Big)  \h_0({\bf y}|{\bf z}_0)  {\rm d} {\bf y} \bigg)\bigg)&\notag\\  \times\Big(\bar{m}_0 \int_{\nbbR^2} v_{0,0}({\bf x},{\bf y})  \h_0({\bf y}|{\bf z}_0)  {\rm d} {\bf y}   +1 \Big)& \notag
\\ \times \prod\limits_{j\in\ncalK\setminus\{0\}}G_j(v_{0,j}({\bf x},{\bf y})) \h_0({\bf x}|{\bf z}_0) f_0({\bf z}_0) \:{\rm d} {\bf x}\: {\rm d} {\bf z}_0,&\text{when $\Phi_0=\ncalB_q^{{\bf z}_0}$ (\caseS~3),}
\end{cases}
\end{align}
where $G_j(\cdot)$ is given by Lemma~\ref{lemm::pgfl_phi_0} and $f_0({\bf z}_0)$ is the  PDF of  ${\bf z}_0$ which is defined in \eqref{eq::phi_0_def}. 
\end{lemma}
\begin{IEEEproof}
\caseC~1 is trivial. For \caseS~2, $\Phi_0$ has only one point  with PDF $f_0({\bf z}_0)$. For \caseS~3, we use Lemma~\ref{lemm::sumproduct::finite} with  $g({\bf x })=\prod\limits_{j\in\ncalK\setminus\{0\}}G_j(v_{0,j}({\bf x},{\bf y}))$ and $v({\bf x},{\bf y})=v_{0,0}({\bf x},{\bf y})$ and take  expectation with respect to ${\bf z}_0\sim f_0({\bf z}_0)$.
\end{IEEEproof}
\subsection{Convergence}\label{sec::convergence}
In this Section, we prove that 
the baseline model 
can be obtained as the limiting case of our general model as cluster size of all the PCPs (i.e. $\Phi_k$, $\forall\ k\in\ncalK_2$ and $\Phi_{\rm u}$ for \caseS~2 and \caseS~3) tends to infinity.
First, we focus on the limiting nature of the BS point process $\Phi'=\cup_{k\in\ncalK_1\cup\ncalK_2} \Phi_{k}$. 
 As the cluster size of  $\Phi_{k}$ $\forall\ k\in\ncalK_2$  increases, the limiting baseline model in this case consists of BS tiers all modeled as PPPs, i.e., $\bar{\Phi}=\cup_{k\in\ncalK_1\cup\ncalK_2} \bar{\Phi}_{k}$, where $\{\bar{\Phi}_k = \Phi_{k}:k\in\ncalK_1\}$ is the collection of the  PPP BS tiers  in the original model and   $\{\bar{\Phi}_{k}:k\in\ncalK_2\}$ is the collection of BS tiers which are also PPP  with intensity $\bar{m}_{k}\lambda_{{\rm p}_{k}}$. We will show that as the cluster size of $\Phi_{k}$ ($k\in\ncalK_2$) goes to infinity, $\Phi_{k}$ converges to $\bar{\Phi}_{k}$ which is independent of the parent PPP $\Phi_{{\rm p}_k}$. 
 
We first formally introduce the notion of increasing the \textit{cluster size} of a PCP $\Phi_k$ ($k\in\ncalK_2$) which means that the points in offspring process (i.e., ${\bf z}+\ncalB^{\bf z}_k$) will lie farther away from the cluster center (${\bf z}\in\Phi_{{\rm p}_k}$) with high probability. One way of modeling this notion is to scale the positions of the offspring points with respect to the cluster center by $\xi$, i.e., ${\bf z}+\ncalB^{\bf z}_k =\{{\bf y}\} = 
\{{\bf z}+\xi{\bf s}\}$. Then the density function defined in $\nbbR^2$ becomes \begin{equation}\label{eq::pcp::density::scaled}
 \h_{k,\xi}({\bf y}|{\bf z})\equiv  f_{k,\xi}({\bf y}-{\bf z})= \frac{1}{\xi^2}f_k\big(\frac{{{\bf y}-{\bf z}}}{\xi}\big), \qquad\forall\ {\bf y}\in {\bf z}+\ncalB_k^{\bf z}.
\end{equation} 
The limiting nature of PCP to PPP is formally proved in the following Proposition. 
\begin{prop}[Weak Convergence of PCP to PPP]
\label{prop::pcp::converge::PPP}
For a PCP $\Phi_{k}\  (\lambda_{{\rm p}_k},f_{k,\xi},\bar{m}_k)$, 
\begin{equation}\label{eq::pcp::convergence::PPP}
\Phi_{k}\to\bar{\Phi}_k\  (weakly)\  \text{as}\ \xi\to\infty,
\end{equation} 
where $\bar{\Phi}_k$ is a PPP of intensity $\bar{m}_k\lambda_{{\rm p}_k}$ if $\sup(f_k)<\infty$.
\end{prop}
\begin{IEEEproof}A simple point process $\Phi_{k}$ ($k\in\ncalK_2$) converges weekly to $\bar{\Phi}_k$ 
if \cite[Theorem 9.1.2]{Mikosch2009}
\begin{subequations}
\begin{equation}\label{eq::convergence::criteria::1}
\nbbE[\Phi_{k}(A)]\to \nbbE[\bar{\Phi}_{k}(A)],
\end{equation}
\begin{equation}\label{eq::convergencve::criteria::2}
\nbbP(\Phi_{k}(A)=0)\to\nbbP(\bar{\Phi}_{k}(A)=0),
\end{equation}
\end{subequations}
for any closed $A\subset \nbbR^2$. Here the same notation has been used to designate a  point process and its associated counting measure. 
Since $\nbbE[\Phi_{k}(A)]=\nbbE[\bar{\Phi}_{k}(A)] = \bar{m}_k\lambda_{{\rm p}_k}$, \eqref{eq::convergence::criteria::1} is satisfied. Next, we observe from \eqref{eq::pcp::density::scaled} that as long as $f_k(\cdot)$ is bounded, 
$f_{k,\xi}({\bf s})\to 0$ as $\xi\to\infty$. 
Now, the void probability of $\Phi_{k}$ i.e. the probability that no points of $\Phi_{k}$ will lie in $A$  along with the limit $\xi\to\infty$ can be written as:
\begin{align*}
&\lim\limits_{\xi\to\infty}\nbbP(\Phi_{k}(A)=0)=\lim\limits_{\xi\to\infty}\nbbE\bigg[\prod\limits_{{\bf z}\in\Phi_{{\rm p}_k}}\prod\limits_{{\bf y}\in{\bf z}+\ncalB^{\bf z}_k}{\bf 1}({\bf  y}\notin A)\bigg]\\
=&\lim\limits_{\xi\to\infty}\exp\bigg(-\lambda_{{\rm p}_k}\int\limits_{\nbbR^2}\bigg(1-\exp\bigg(-\bar{m}_k\bigg(1-\int\limits_{\nbbR^2\setminus A} f_{k,\xi}({\bf y}-{\bf z}){\rm d}{\bf y}\bigg)\bigg)\bigg){\rm d}{\bf z}\bigg)\\
=& \lim\limits_{\xi\to\infty}\exp\bigg(-\lambda_{{\rm p}_k}\int\limits_{\nbbR^2}\bigg(1-\exp\bigg(-\bar{m}_k
\int\limits_{A} f_{k,\xi}({\bf y}-{\bf z})
{\rm d}{\bf y}\bigg)\bigg){\rm d}{\bf z}\bigg)\\
\myeq{a}&\lim\limits_{\xi\to\infty}\exp\bigg(-\lambda_{{\rm p}_k}\bar{m}_k\int\limits_{\nbbR^2}
\int\limits_{A} f_{k,\xi}({\bf y} - {\bf z})
{\rm d}{\bf y}\:{\rm d}{\bf z}\bigg)\myeq{b} \exp\bigg(-\lambda_{{\rm p}_k}\bar{m}_k|A|\bigg)=\nbbP(\bar{\Phi}_k(A)= 0),
\end{align*}
where (a) follows from Taylor series expansion of the exponential function under integration and neglecting the higher order terms as $\xi\to\infty$ and (b) follows from interchanging the order of integrals and the fact that  $|A|$ is finite. 
\end{IEEEproof}
We now argue that as $\xi\to\infty$, $\Phi_{k}$ becomes independent of its parent PPP $\Phi_{{\rm p}_k}$.  
\begin{prop}\label{prop::PCP::PPP::Convergence::independence}
The limiting PPP $\bar{\Phi}_k$ and the parent PPP $\Phi_{{\rm p}_k}$ of  $\Phi_{k}$ ($k\in\ncalK_2$) are independent, i.e.,
\begin{align}
\lim\limits_{\xi\to\infty}\nbbP(\Phi_{k}(A_1) = 0, \Phi_{{\rm p}_k}(A_2)=0) = \nbbP(\bar{\Phi}_{k}(A_1)=0)\nbbP(\Phi_{{\rm p}_k}(A_2)= 0),\label{eq::PCP::PPP::Convergence::independence}
\end{align} 
where  $A_1, A_2\subset \nbbR^2$ are arbitrary closed compact sets. 
\end{prop}
\begin{IEEEproof} Following Choquet theorem for random closed sets \cite[Theorem 6.1]{chiu2013stochastic}, \eqref{eq::PCP::PPP::Convergence::independence} is a sufficient condition to claim independence of $\bar{\Phi}_k$ and $\Phi_{{\rm p}_k}$. 
Under the limit $\xi\to \infty$: 
\begin{align*}
&\lim\limits_{\xi\to\infty}\nbbP(\Phi_{k}(A_1)=0, \Phi_{{\rm p}_k}(A_2)=0)=  \lim\limits_{\xi\to\infty}\nbbP(\Phi_{k}(A_1)=0| \Phi_{{\rm p}_k}(A_2)=0)\nbbP(\Phi_{{\rm p}_k}(A_2)=0)\\
=&\lim\limits_{\xi\to\infty}\nbbE\bigg[\prod\limits_{{\bf z}\in\Phi_{{\rm p}_k}\cap A_2^c}\prod\limits_{{\bf y}\in{\bf z}+\ncalB^{\bf z}_k}{\bf 1}({\bf  y}\notin A_1)\bigg]\nbbP(\Phi_{{\rm p}_k}(A_2)=0)\\
=& \lim\limits_{\xi\to\infty}\exp\bigg(-\lambda_{{\rm p}_k}\int\limits_{\nbbR^2\setminus A_2}\bigg(1-\exp\bigg(-\bar{m}_k
\int\limits_{A_1} f_{k,\xi}({\bf y}-{\bf z})
{\rm d}{\bf y}\bigg)\bigg){\rm d}{\bf z}\bigg)\nbbP(\Phi_{{\rm p}_k}(A_2)=0)\\
\myeq{a}&\lim\limits_{\xi\to\infty}\exp\bigg(-\lambda_{{\rm p}_k}\bar{m}_k\int\limits_{\nbbR^2\setminus A_2}
\int\limits_{A_1} f_{k,\xi}({\bf y}-{\bf z})
{\rm d}{\bf y}{\rm d}{\bf z}\bigg)\nbbP(\Phi_{{\rm p}_k}(A_2)=0)\\ 
 =& \lim\limits_{\xi\to\infty}\exp\bigg(-\lambda_{{\rm p}_k}\bar{m}_k\int\limits_{\nbbR^2}
\int\limits_{A_1} f_{k,\xi}({\bf y} - {\bf z})
{\rm d}{\bf y}\:{\rm d}{\bf z}\bigg)\exp\bigg(\lambda_{{\rm p}_k}\bar{m}_k\int\limits_{A_2}
\int\limits_{A_1} f_{k,\xi}({\bf y} - {\bf z})
{\rm d}{\bf y}\:{\rm d}{\bf z}\bigg) \nbbP(\Phi_{{\rm p}_k}(A_2)=0)\\
\myeq{b}&
 \exp\bigg(-\lambda_{{\rm p}_k}\bar{m}_k|A_1|\bigg)\nbbP({\Phi}_{{\rm p}_k}(A_2)= 0) = \lim\limits_{\xi\to\infty} \nbbP(\Phi_{k}(A_1) = 0)\nbbP({\Phi}_{{\rm p}_k}(A_2)= 0),
\end{align*}
  where (a) follows on the similar lines of step (a) in the proof of Proposition~\ref{prop::pcp::converge::PPP}. In (b), we apply the limit $\xi\to\infty$. The first term in the product follows from  Proposition~\ref{prop::pcp::converge::PPP} and the second term goes to $1$ as the double integral over a finite region ($A_1\times A_2$) tends to zero as $\lim_{\xi\to 0 } f_{k,\xi}({\bf s})=0$. 
\end{IEEEproof}
\begin{remark}
   Using Propositions~\ref{prop::pcp::converge::PPP} and \ref{prop::PCP::PPP::Convergence::independence},  we can claim that  the $K$-tier HetNet model under \caseS~2 ($\Phi_{\rm u}$ is a PCP around $\Phi_q$ ($q\in\ncalK_1$)) converges to that of \caseS~1 (i.e., users form  a  PPP independent of BS locations) as the cluster size of $\Phi_{\rm u}$  increases to infinity. 
 Further, for \caseS~3, where $\Phi_{\rm u}$ and $\Phi_{q}$  are coupled by the same parent PPP $\Phi_{{\rm p}_q}$, as the cluster size of $\Phi_{\rm u}$ as well as    $\Phi_q$ ($q\in\ncalK_2$) increase  to infinity,  
   $\Phi_{\rm u}$ and $\Phi_{\rm q}$ become independent PPPs.  
\end{remark}
From this Proposition, we can directly conclude the following.
\begin{cor}When cluster size of $\Phi_k$, $\forall\ k\in\ncalK_2$ tends to infinity,  coverage probability  
can be written as \cite[Corollary 1]{dhillon2012modeling}:
\begin{align}
\pc = \frac{\pi}{C(\alpha)}\frac{\sum_{k\in\ncalK_1}\frac{\lambda_kP_k^{\frac{2}{\alpha}}}{\beta_k^{\frac{2}{\alpha}}}+\sum_{k\in\ncalK_2}\frac{\bar{m}_k\lambda_{{\rm p}_k}P_k^{\frac{2}{\alpha}}}{\beta_k^{\frac{2}{\alpha}}}}{\sum_{j\in\ncalK_1}\lambda_j P_j^{\frac{2}{\alpha}}+\sum_{j\in\ncalK_2}\bar{m}_j\lambda_{{\rm p}_j} P_j^{\frac{2}{\alpha}}},
\end{align}
where   $C(\alpha) = \frac{\alpha}{2 \pi} {\sin(\frac{2 \pi} {\alpha})}$.
\end{cor}
 \begin{figure}
 \begin{minipage}{.48\textwidth}
 \centering
 \includegraphics[width =0.6\textwidth]{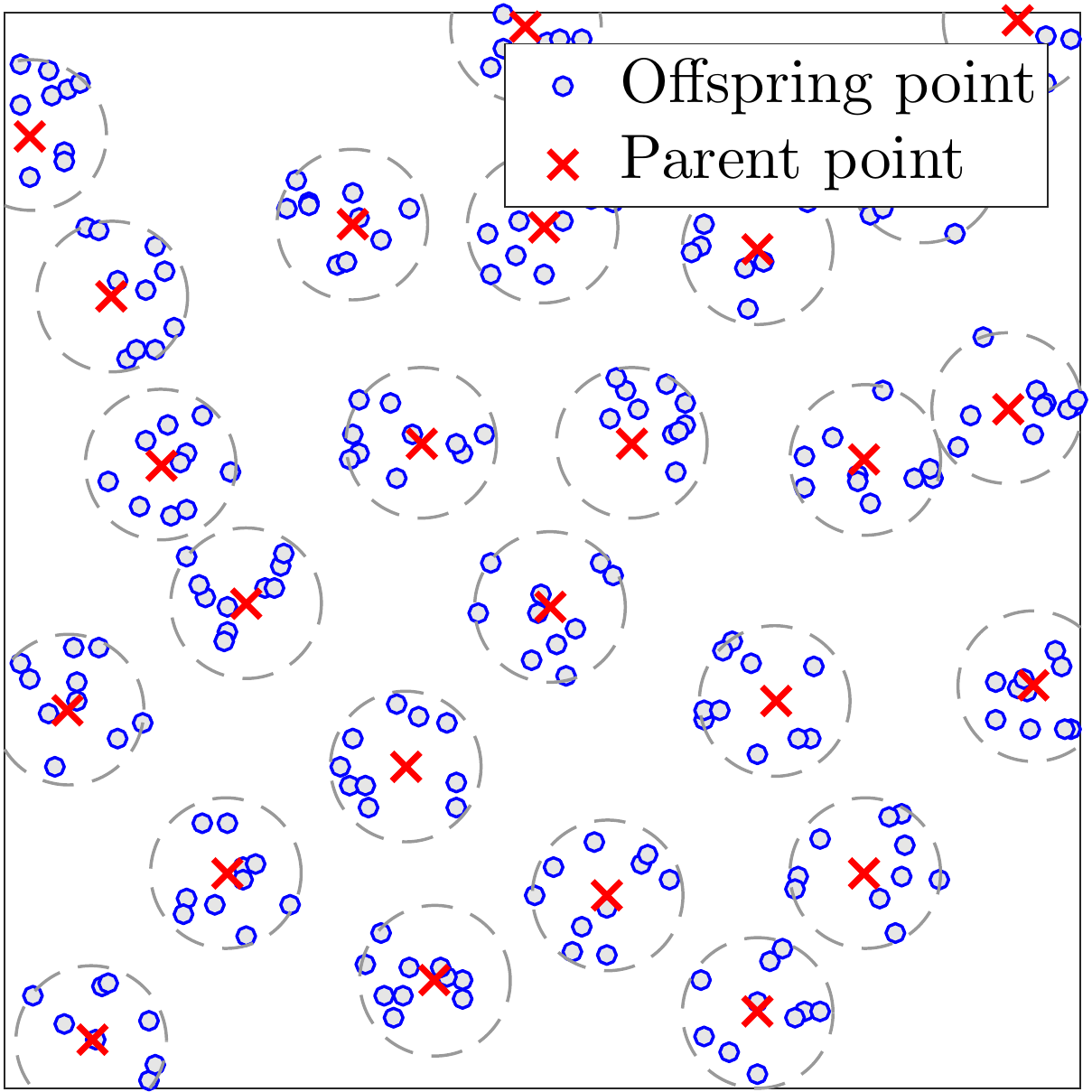}
\caption{\small A realization of a \matern cluster process.}\label{fig::matern}
\end{minipage}
\hfill
 \begin{minipage}{.48\textwidth}
 \centering
 \includegraphics[width =0.6\textwidth]{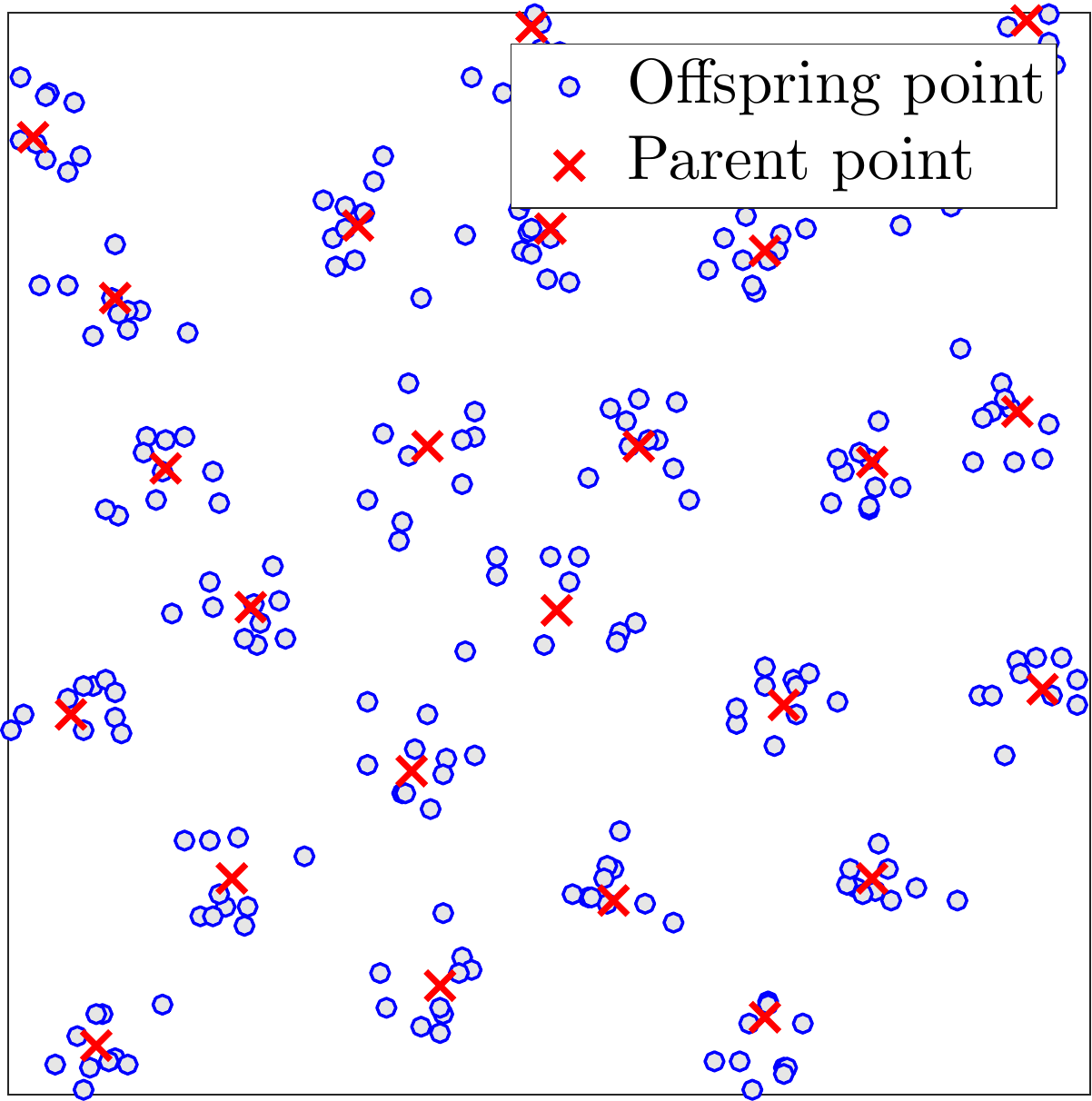}
\caption{\small A realization of a Thomas cluster process.} \label{fig::thomas}
\end{minipage}
\end{figure}
Having derived the expression for coverage probability under the general framework,  we now focus on two special cases as follows.
\subsection{\matern Cluster Process}
We assume that all BS tiers $\Phi_k$, $\forall\ k\in\ncalK_2$ and  user tier $\Phi_{\rm u}$ (for \caseS~2 and \caseS~3) are modeled as MCP.  We choose MCP for $\Phi_k$ ($\forall\ k\in\ncalK_2$)  since it closely resembles 3GPP model for SBS and user clusters. We first 
formally define MCP $\Phi_k$ ($k\in\ncalK_2$) as follows. 
\begin{ndef}[MCP] A PCP  $\Phi_k\ (\lambda_{{\rm p}_k},f_k,\bar{m}_k)$ is called a MCP if the distribution of the offspring points in $\ncalB_k^{\bf z}$ is uniform within a disc of radius $r_{{\rm d}_k}$ around the origin denoted by $b({\bf 0},r_{{\rm d}_k})$, i.e., if ${\bf s}=(\|{\bf s}\|,\arg({\bf s}))\equiv(s,\theta_s)\in\ncalB_k^{\bf z}$ denotes a point of the offspring point process ${\ncalB_k^{\bf z}}$ with cluster center at origin, then the joint PDF of the polar coordinates of ${\bf s}$ is denoted by:
\begin{align}\label{eq::density_matern_definition}
f_k({\bf s}) =f_k(s,\theta_s) = \frac{2s}{r_{{\rm d}_k}^2}{\times\frac{1}{2\pi}}, \qquad 0<s\leq r_{{\rm d}_k},0< \theta_s\leq 2\pi.
\end{align}  
\end{ndef} 
Note that we will use $(s,\theta_s)$ and $(\|{\bf s}\|,\arg({\bf s}))$ as  the  representation of ${\bf s}\in\nbbR^2$  
in Polar coordinates interchangeably. 
A realization of an MCP is illustrated in \figref{fig::matern}. 
 First, we observe that the functions associated with the sum-product functional in the coverage probability expression in Theorem~\ref{thm::coverage} are isotropic, i.e., referring to \eqref{eq::sum::product::defn},   $v({\bf x},{\bf y})=v(x,y)\equiv v_{k,k}(x,y)$ and $g({\bf x})=g(x)\equiv \prod_{j\in\ncalK\setminus\{k\}}G_j(v_{k,j}(x,y))$, $\forall\ k,j\in\ncalK$. Thus, the sum-product functional for $\Phi_k$ appearing in $\pc_k$ in \eqref{eq::coverage_main_result} is in the form: $\nbbE\sum_{{\bf x}\in\Phi_k}g(x)\prod_{{\bf y}\in\Phi_k\setminus{\{\bf x}\}}v(x,y)$. { Following Lemmas~\ref{lemm::sumproduct::ppp},  \ref{lemm::sumproduct::pcp} and \ref{lemm::pgfl::cluster}, it is sufficient to evaluate the PGFLs $G_j(v_{k,j}(x,y))$ and $G_{c_j}(v_{k,j}(x,y))$ for $\pc_k$, which we do next.} We will use these results to derive  the final expression  of coverage probability.
 
\begin{remark}\label{rem::marginal::distribution}We  observe that the integrals appearing in \eqref{eq::pgfl::pcp} and \eqref{eq::coverage_main_result} are in the form:
$$\int_{0}^{\infty}\int_{0}^{2\pi}\rho({ x},{ z})\h_k({\bf x}|{\bf z}){\rm d}{ x}\:{\rm d}{\theta_x}=\int_{0}^{\infty}\rho({ x},{ z})\int_{0}^{2\pi}\h_k({x},\theta_x|{\bf z}){\rm d}{\theta_x}\:{\rm d}{ x}.$$Here $\int_{0}^{2\pi}\h_k({ x},\theta_x|{\bf z}){\rm d}{\theta_x}$ is the marginal distribution of the magnitude of ${\bf x}\in\Phi_{k}$ ($k\in\ncalK_2$) conditioned on ${\bf z}\in\Phi_{{\rm p}_k}$.
\end{remark}
In order to  characterize  the conditional magnitude distribution of ${\bf x}$ given ${\bf z}\in \Phi_{{\rm p}_k}$,  we  define three regions $\ncalR_k^{(1)}, \ncalR_k^{(2)}, \ncalR_k^{(3)}\subset \nbbR^2\times\nbbR^2$ as:
\begin{subequations}
\begin{alignat}{4}
&\ncalR_k^{(1)} \equiv {\bf z} \in b({\bf 0},r_{{\rm d}_k}),\ {\bf x}\in b({\bf 0}, r_{{\rm d}_k}-z),\\
&\ncalR_k^{(2)} \equiv {\bf z} \in b({\bf 0},r_{{\rm d}_k}), \ {\bf x}\in b({\bf z},r_{{\rm d}_k})\setminus b({\bf 0}, r_{{\rm d}_k}-z),\\
&\ncalR_k^{(3)} \equiv {\bf z} \notin b({\bf 0},r_{{\rm d}_k}), \  {\bf x}\in b({\bf z},r_{{\rm d}_k}). 
\end{alignat}
\end{subequations}
\begin{figure*}
\begin{subfigure}{0.5\textwidth}
\centering
\includegraphics[width =0.6\textwidth]{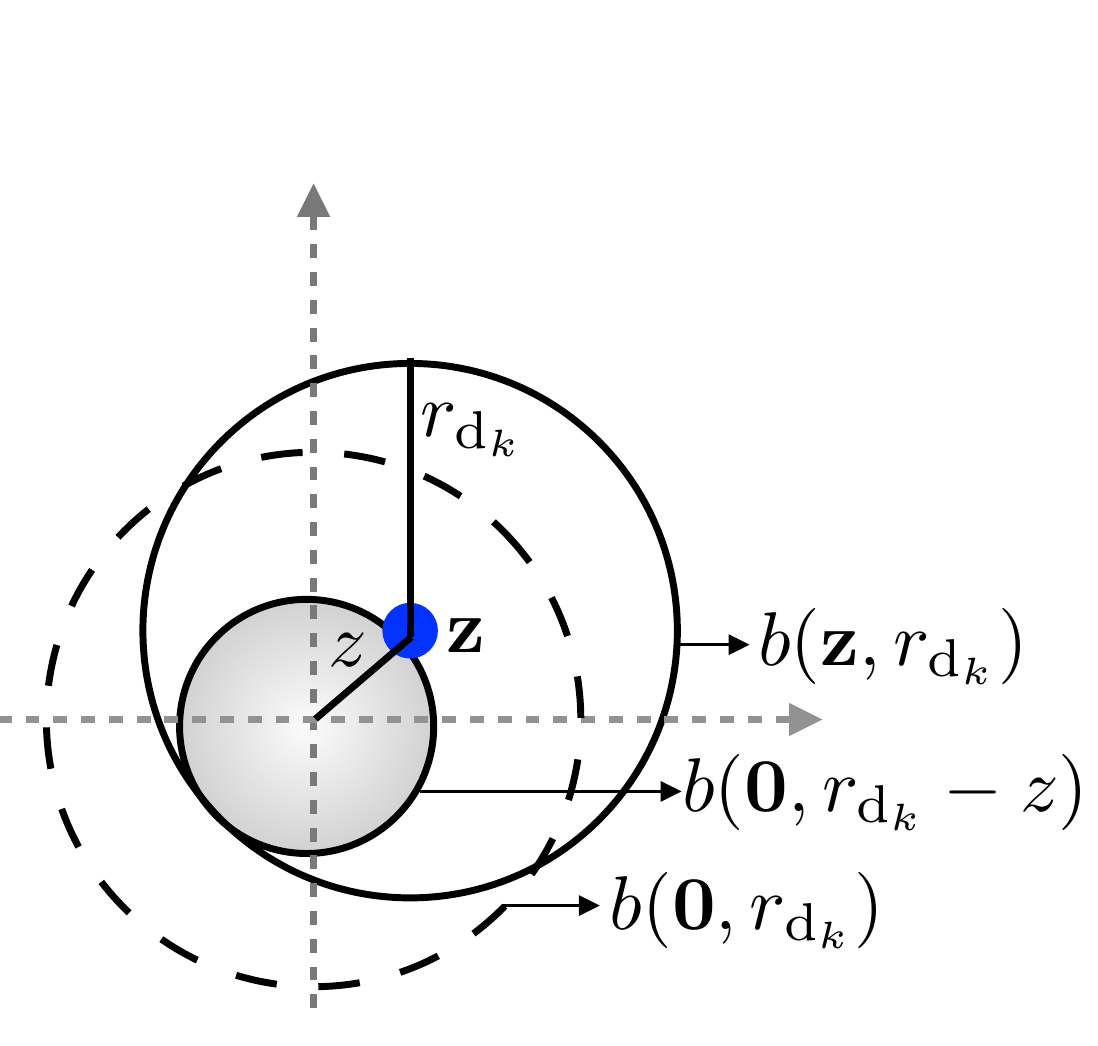}
\caption{${\bf z} \in b({\bf 0},r_{{\rm d}_k}),\ {\bf x}\in b({\bf 0}, r_{{\rm d}_k}-z)$, or\\ ${\bf z} \in b({\bf 0},r_{{\rm d}_k}), \ {\bf x}\in b({\bf z},r_{{\rm d}_k})\setminus b({\bf 0}, r_{{\rm d}_k}-z)$.
}\label{fig::matern::construction::case1}
\end{subfigure}
\begin{subfigure}{0.5\textwidth}
\centering
\includegraphics[width = 0.6\textwidth]{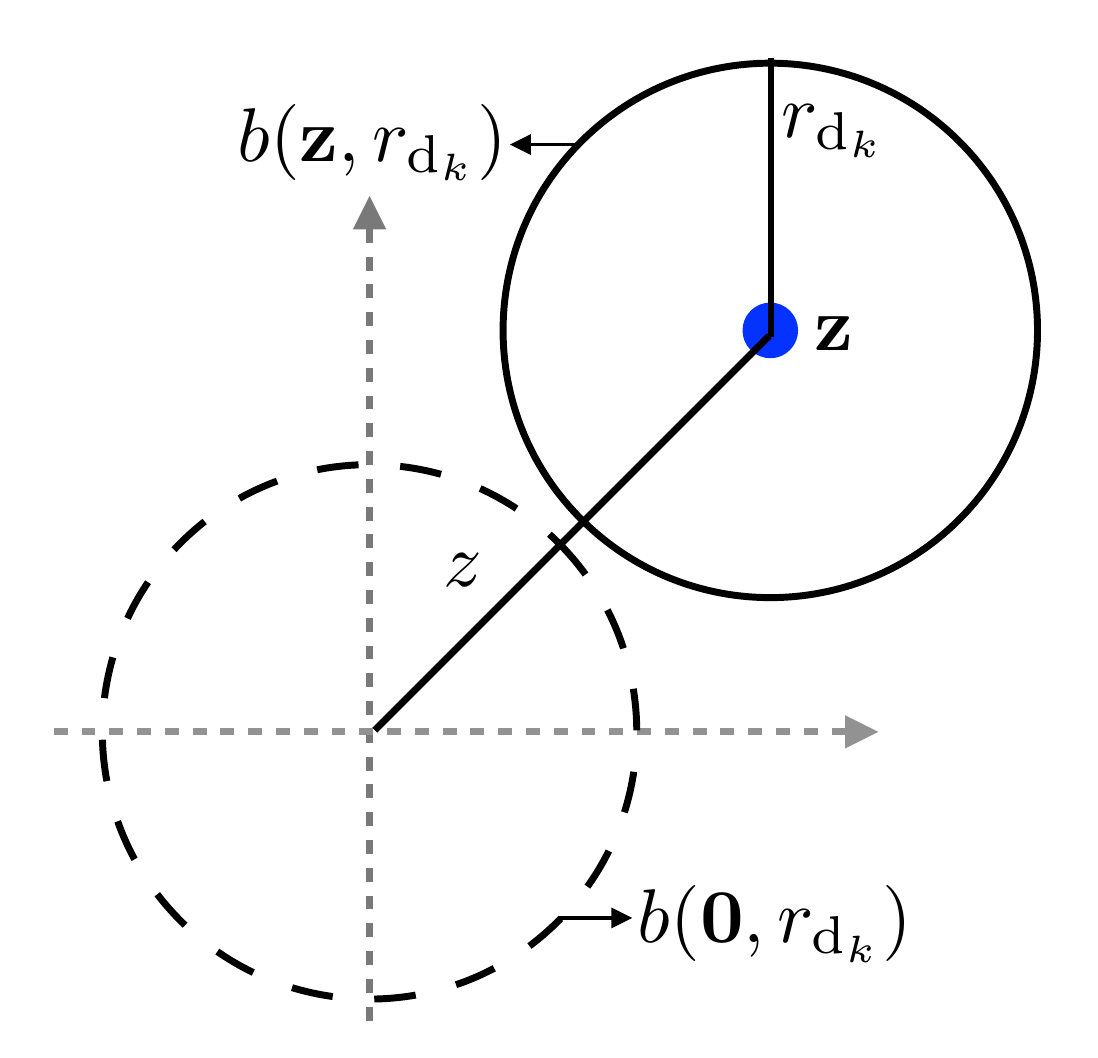}
\caption{$ {\bf z} \notin b({\bf 0},r_{{\rm d}_k}), \  {\bf x}\in b({\bf z},r_{{\rm d}_k}). $}\label{fig::matern::construction::case2}
\end{subfigure}
\caption{\small Possible positions of a cluster center at ${\bf z}$ for the evaluation of the  distribution of distance of a randomly chosen point ${\bf x}\in{\bf z}+\ncalB^{{\bf z}}_k$ of an MCP from origin.}
\label{fig::matern::construction}
\end{figure*}
Illustrations of these regions are provided in \figref{fig::matern::construction}. For each region, the marginal distribution of $x$ conditioned on $\bf z$ is given by~\cite{ZhuGovindasamyHwang2014}: $\int_{0}^{2\pi}\h_k({ x},\theta_x|{\bf z}){\rm d}{\theta_x}=
\chi_k^{(\ell)}(x,z)\text{ when }({\bf z},{\bf x})\in \ncalR_k^{(\ell)}\ (\ell=1,2,3)$,
 where 
\begin{subequations}\label{eq::chi_definition}
\begin{alignat}{4}
&\chi_k^{(1)}(x,z) = \frac{2 x}{r_{{\rm d}_k}^2},& 0<x<r_{{\rm d}_k}-z, 0<z\leq r_{{\rm d}_k},\\
&\chi_k^{(2)}(x,z) = \frac{2 x}{\pi r_{{\rm d}_k}^2}\cos^{-1}\left(\frac{x^2+z^2-r_{{\rm d}_k}^2}{2xz}\right), & r_{{\rm d}_k}-z<x<r_{{\rm d}_k}+z, 0<z\leq r_{{\rm d}_k},\\
&\chi_k^{(3)}(x,z) = \frac{2 x}{\pi r_{{\rm d}_k}^2}\cos^{-1}\left(\frac{x^2+z^2-r_{{\rm d}_k}^2}{2 x z}\right),&z-r_{{\rm d}_k}<x<z+{r}_{{\rm d}_k}, z>r_{{\rm d}_k}.
\end{alignat}
\end{subequations}
We now present the expressions of PGFLs of $\Phi_k$ ($k\in\ncalK_2$).
\begin{cor}[PGFL of MCP] \label{cor::pgfl::mcp}
The PGFL of MCP $\Phi_j$ ($j\in\ncalK_2$) evaluated at $v_{k,j}(x,y)$ is: 
\begin{multline}\label{eq::pgfl::mcp}
{G}_j(v_{k,j}(x,{y}))=\exp\bigg(-2\pi\lambda_{{\rm p}_j}\int\limits_{0}^{r_{{\rm d}_j}}\bigg(1-\exp\bigg(-\bar{m}_j \bigg(\int\limits_{0}^{r_{{\rm d}_j}-z} (1-v_{k,j}(x,y))\chi_j^{(1)}(y,z){\rm d}{y}\\
+\int\limits_{r_{{\rm d}_j}-z}^{r_{{\rm d}_j}+z}(1-v_{k,j}(x,y))\chi_j^{(2)}(y,z){\rm d}{ y}\bigg) \bigg)\bigg)z{\rm d}{z}\\
-2\pi\lambda_{{\rm p}_j}\int\limits_{r_{{\rm d}_j}}^{\infty}\bigg(1-\exp\bigg(- \bar{m}_j\int\limits_{z-r_{{\rm d}_j}}^{z+r_{{\rm d}_j}}(1-v_{k,j}(x,y)\chi_j^{(3)}(y,z){\rm d}{y}\bigg)\bigg)  z{\rm d}{z}\bigg),
\end{multline}
where $\chi_j^{(\ell)}(x,z)$ ($\ell=1,2,3$) are given by \eqref{eq::chi_definition}. 
\end{cor}
\begin{IEEEproof}
The expression can be derived from \eqref{eq::pgfl::pcp} using Remark~\ref{rem::marginal::distribution}.
\end{IEEEproof}
\begin{cor}[PGFL of Offspring Point Process of MCP]
\label{cor::pgfl::offspring::mcp}
The PGFL of  $\ncalB_j^{\bf z}$, which is the offspring process of $\Phi_j$ ($j\in\ncalK_2$) centered at ${\bf z}$, evaluated at $v_{k,j}(x,y)$ is given by 
\begin{align}\label{eq::pgfl::mcp::reduced::cluster}
{G}_{c_j}(v_{k,j}(x,y)|{\bf z})=\begin{cases}{G}_{c_j}^{(1)}(v_{k,j}(x,y)|{\bf z})& z\leq r_{{\rm d}_j}\\
{G}_{c_j}^{(2)}(v_{k,j}(x,y)|{\bf z}) & z>r_{{\rm d}_j},
\end{cases}
\end{align}
where 
${G}_{c_j}^{(1)}(v_{k,j}(x,y)|{\bf z})\equiv \exp\bigg( -\bar{m}_j\bigg(\int\limits_{0}^{r_{{\rm d}_j}-z} (1-v_{k,j}(x,y)) \chi^{(1)}_j(y,z){\rm d}{y}
+\int\limits_{r_{{\rm d}_j}-z}^{r_{{\rm d}_j}+z}v_{k,j}(x,y) \chi_j^{(2)}(y,z){\rm d}{y}\bigg)\bigg)$ and  ${G}_{c_j}^{(2)}(v_{k,j}(x,y)|{\bf z})\equiv  \exp\bigg(-\bar{m}_j\int\limits_{z-r_{{\rm d}_j}}^{z+r_{{\rm d}_j}}(1-v_{k,j}({x},{y})) \chi_{j}^{(3)}(y,z){\rm d}{y}\bigg)$.

\end{cor}
\begin{IEEEproof}
The expression can be derived from \eqref{eq::pgfl::reduced::cluster} using Remark~\ref{rem::marginal::distribution}.
\end{IEEEproof}
We can now obtain the PGFL of an MCP under its reduced Palm distribution at $v_{k,k}(x,y)$ by rewriting \eqref{eq::reduced::palm::pcp} and using \eqref{eq::pgfl_cluster_reduced} as: 
\begin{align}
\widetilde{G}_k(v_{k,k}(x,y)|{\bf z}) = \begin{cases}\widetilde{G}_k^{(1)}(v_{k,k}(x,y)|{\bf z}) \equiv G_k(v_{k,k}(x,y)){G}_{c_k}^{(1)}(v_{k,k}(x,y)|{\bf z}),&z\leq r_{{\rm d}_k},\\
\widetilde{G}_k^{(2)}(v_{k,k}(x,y)|{\bf z}) \equiv G_k(v_{k,k}(x,y)){G}_{c_k}^{(2)}(v_{k,k}(x,y)|{\bf z}),&z>r_{{\rm d}_k},
\end{cases}
\end{align}
 where $G_k(v_{k,k}(x,y))$ and $\widetilde{G}_{c_k}(v_{k,k}(x,y)|{\bf z})$ are given by Corollaries~\ref{cor::pgfl::mcp} and \ref{cor::pgfl::offspring::mcp}, respectively. 
We are left with the  PGFL of $\Phi_0$, i.e., $G_{0}(v_{0,k}(x,y))$  can be obtained by substitution of {$f_0(\cdot)$ } in Lemma~\ref{lemm::pgfl_phi_0} with \eqref{eq::density_matern_definition}. For \caseS~2, this can be given as:
\begin{align*}
{G_0(v_{k,0}({x},{ y})) = \int\limits_{0}^{r_{{\rm d}_0}}\frac{1}{1+\frac{P_0\beta_k}{P_k}\big(\frac{x}{y}\big)^{-\alpha}}\frac{2y}{r_{{\rm d}_0}^2}{\rm d}y.}
\end{align*}
For \caseS~3, 
\begin{align*}
G_0(v_{k,0}({ x},{ y})) = \int\limits_{0}^{r_{{\rm d}_0}}G_{c_0}(v_{k,0}(x,y)|z_0)\frac{2z_0}{r_{{\rm d}_0}^2}{\rm d}z_0.
\end{align*}
We now present the expression of per-tier coverage $\pc_k$ for $k\in\ncalK_2$.  
\begin{cor}\label{corr::coverage::mcp}Per-tier coverage probability for $k\in\ncalK_2$ when all BS tiers in $\ncalK_2$ are modeled as MCPs can be expressed as:
\begin{multline*}
\pc_k =  2 \pi\lambda_{{\rm p}_k}\bar{m}_k\int\limits_{0}^{r_{{\rm d}_k}}\int\limits_{0}^{r_{{\rm d}_k}-z}g(x){G}_k(v_{k,k}(x,y)){G}^{(1)}_{c_k}(v_{k,k}(x,y)|{\bf z})\chi_k^{(1)}(x,z){\rm d}x\: z {\rm d}z\\+ 2 \pi\lambda_{{\rm p}_k}\bar{m}_k\int\limits_{0}^{r_{{\rm d}_k}}\int\limits_{r_{{\rm d}_k}-z}^{r_{{\rm d}_k}+z}g(x){G}_k(v_{k,k}(x,y)){G}^{(1)}_{c_k}(v_{k,k}(x,y)|{\bf z})\chi_k^{(2)}(x,z){\rm d}x\: z{\rm d}z\\+2 \pi \lambda_{\rm p}\bar{m}_k\int\limits_{r_{{\rm d}_k}}^{\infty}\int\limits_{z-r_{{\rm d}_k}}^{z-r_{{\rm d}_k}}g(x)G_k(v_{k,k}(x,y)){G}^{(2)}_{c_k}(v_{k,k}(x,y)|{\bf z})\chi_k^{(3)}(x,z){\rm d}x\: z {\rm d}z, \ k\in\ncalK_2,
\end{multline*}
where $g(x) =\prod\limits_{j\in\ncalK_1}G_j(v_{k,j}(x,y)) \prod\limits_{j\in\ncalK_2\setminus\{k\}}G_j(v_{k,j}(x,y))$. Here $G_j(v_{k,j}(x,y))$ is given by \eqref{eq::pgfl_ppp_applied} and  \eqref{eq::pgfl::mcp}  for $j\in\ncalK_1$ and $j\in\ncalK_2$, respectively, and ${G}_{c_k}^{(1)}(\cdot)$, ${G}_{c_k}^{(2)}(\cdot)$ are given by Corollary~\ref{cor::pgfl::offspring::mcp}.  
\end{cor}
\begin{IEEEproof}
The expression is obtained from Lemma~\ref{lem:per-tier-cov-pcp} by using the Polar domain representation of the vectors and the distance distribution introduced in \eqref{eq::chi_definition}. 
\end{IEEEproof}
As noted earlier, $\pc_0$ can be obtained by computing sum-product functional over $\Phi_0$ which has three different forms depending on the user configuration. While \caseS~1 and \caseS~2 are simple, for  \caseS~3, we  need to evaluate sum-product functional of ${\bf z}+\ncalB^{\bf z}_k$. 
\begin{cor}\label{corr::pc_0_mcp}Per-tier coverage probability for $k=0$ when all BS tiers in $\ncalK_2$ are modeled as MCPs can be expressed as:
\begin{align*}
\pc_0 =  \begin{cases}
 0,&\text{\caseS~1},\\
 \int\limits_{0}^{r_{{\rm d}_0}}\prod\limits_{j\in\ncalK\setminus\{0\}}G_j(v_{0,j}({z_0},{y}))\bar{f}_0({z_0}){\rm d}{z_0} ,&\text{\caseS~2},\\ 
\int\limits_{0}^{r_{{\rm d}_0}} \int_{0}^{{r_{{\rm d}_0}}-z_0}\big[\ncalH(x,z_0)\chi_0^{(1)}({x},{z_0}){\rm d}{x}  +\int_{{r_{{\rm d}_0}}-z_0}^{{r_{{\rm d}_0}}+z_0} \ncalH(x,z_0)\chi_0^{(2)}(x,z_0){\rm d}{x} \big]     2z_0/r_{{\rm d}_0}^2{\rm d}z_0,&\text{\caseS~3},
  \end{cases}
\end{align*}
and
\begin{multline}
\ncalH{(x,z)}=\prod\limits_{j\in \ncalK\setminus\{0\}}G_j(v_{0,j}(x,y))\exp\bigg(-\bar{m}_0\Big( \int_0^{r_{{\rm d}_0} -z} (1-v_{0,0}({ x},{ y})) \chi^{(1)}_0(y,z) {\rm d}{ y}\\+ \int_{r_{{\rm d}_0} -z}^{r_{{\rm d}_0 }+z} (1-v_{0,0}({ x},{ y})) \chi_0^{(2)}(y,z) {\rm d}{ y}\Big)\bigg)\bigg( \bar{m}_0\Big(\int_{0}^{r_{{\rm d}_0} -z} v_{0,0}(x, y)  \chi^{(1)}_0(y,z) {\rm d}{ y}\\ +\int_{0}^{r_{{\rm d}_0} -z} v_{0,0}(x, y)  \chi_0^{(2)}(y,z) {\rm d}{ y} \Big) +1  \bigg).
\end{multline}
\end{cor}
\begin{IEEEproof}
For \caseS~1  and  \caseS~2, the result follows directly from Lemma~\ref{lemm::tier0}. For  \caseS~3, we need the sum-product functional of $\Phi_0 = {\bf z}_0 +\ncalB^{{\bf z}_0}_q\equiv {\bf z}_0 +\ncalB^{{\bf z}_0}_0$. Now, by construction, 
$z_0< r_{{\rm d}_q}\equiv r_{{\rm d}_0}$. Since the representative BS cluster $\ncalB^{{\bf z}_0}_0$ has the same cluster center ${\bf z}_0 $ of the typical user located at origin. 
We first evaluate the sum-product functional of ${\bf z}+\ncalB_0^{{\bf z}}$  following Lemma~\ref{lemm::sumproduct::finite}, which can be written as: $\nbbE\Big[\sum\limits_{{\bf x}\in ({\bf z}+\ncalB^{{\bf z}}_0)}g(x)\prod\limits_{{\bf y}\in({\bf z}+\ncalB_0^{{\bf z}})\setminus\{{\bf x}\}}v(x,y)\Bigg]=$
\begin{align*}
g(x)\exp\bigg(-\bar{m}_0\Big( \int_0^{r_{{\rm d}_0} -z} (1-v(x,{ y})) \chi^{(1)}_0(y,z) {\rm d}{ y}+ \int_{r_{{\rm d}_0} -z}^{r_{{\rm d}_0 }+z} (1-v({ x},{ y})) \chi_0^{(2)}(y,z) {\rm d}{ y}\Big)\bigg)\\\times\bigg( \bar{m}_0\Big(\int_{0}^{r_{{\rm d}_0} -z} v(x, y)  \chi^{(1)}_0(y,z) {\rm d}{ y} +\int_{r_{{\rm d}_0} -z}^{r_{{\rm d}_0} +z} v(x, y)  \chi_0^{(2)}(y,z) {\rm d}{ y} \Big) +1  \bigg),\quad z\leq r_{{\rm d}_0}.
\end{align*}
Now substituting $g(x)$ by $\prod\limits_{j\in\ncalK\setminus\{0\}}G_j(v_{0,j}(x,y))$ and $v(x,y)$ by $v_{0,0}(x,y)$ (given by \eqref{eq::coverage_main_result} and \eqref{eq::v_function}, respectively) and deconditioning over $z_0$, we get the final form.\end{IEEEproof}

\subsection{Thomas Cluster Process}
We further provide the results of coverage probability when all BS tiers $\Phi_k$, $\forall\ k\in\ncalK_2$ are modeled as TCP. We first formally define TCP as follows. 
\begin{ndef}[TCP]\label{def::TCP}
 A PCP $\Phi_k\ (\lambda_{{\rm p}_k},f_k,\bar{m}_k)$ is called a  TCP if the distribution of the offspring points in $\ncalB_k^{\bf z}$ is Gaussian around the cluster center at origin, i.e. for all ${\bf s}\in\ncalB_k^{\bf z}$,
\begin{align}\label{eq::density_thomas_definition}
f_k({\bf s}) =f_k( {s},\theta_s)=  \frac{s}{\sigma_k^2}\exp\left(-\frac{s^2}{2\sigma_k^2}\right)\frac{1}{2\pi},\ s>0,  0<\theta_s\leq 2\pi.
\end{align}  
\end{ndef}
A realization of a TCP is illustrated in \figref{fig::thomas}. 
 It will be evident at the end of this Section that compared to MCP, TCP yields simpler expression of coverage probability (due to infinite support of $f_k({\bf s})$). Note that while TPC does not directly analogous to the notion of {\em cluster} adopted in 3GPP HetNet, we include it here to demonstrate the generality of the proposed framework that surpasses that of the cluster-based simulation models adopted by 3GPP. 
Given that ${\bf z}$ is the cluster center of $
{\bf x}$, i.e., ${\bf x}\in{\bf z}+{\cal B}_k^{\bf z}$, we write the conditional PDF of $x$
 as \cite{AfshDhi2015MehrnazD2D1}:
 \begin{align}\label{eq::marginal::dist::tcp}
\int_{0}^{2\pi}\h_k(x,\theta_x|{\bf z}) {\rm d}{\theta_x}= \Omega_k(x,z)= \frac{x}{ \sigma_k^2} \exp\left(-\frac{x^2+z^2}{2 \sigma_k^2}\right) I_0\left(\frac{x z}{\sigma_k^2}\right), \qquad x,z>0.
\end{align}
As we have done for MCP, we first provide the expressions of $G_j(v_{k,j}(x,y))$ and $G_{c_j}(v_{k,j}(x,y))$  for $j\in\ncalK_2$. 
\begin{cor}[PGFL of TCP] \label{cor::pgfl::tcp}
The PGFL of TCP  $\Phi_j$ evaluated at $v_{k,j}(x,y)$ is given by:
\begin{equation*}
{G}_j(v_{k,j}({x},{y}))=\exp\bigg(-2\pi\lambda_{{\rm p}_j}\int\limits_{0}^{\infty}\bigg(1-\exp\bigg(1-\bar{m}_j \bigg(\int\limits_{0}^{\infty} (1-v_{k,j}({x},{y}))\Omega_j(y,z){\rm d}{y}\bigg)\bigg)z\:{\rm d}{z}\bigg).
\end{equation*}
\end{cor}
\begin{IEEEproof} Similar to Corollary~\ref{cor::pgfl::mcp},   the expression can be derived from \eqref{eq::pgfl::pcp} using Remark~\ref{rem::marginal::distribution}.
\end{IEEEproof}
\begin{cor}[PGFL of Offspring Point Process of TCP]
\label{cor::pgfl::offspring::tcp}
 When ${\bf z}+\ncalB_j^{\bf z}$ is the offspring process of a TCP $\Phi_j$, its PGFL   evaluated at $v_{k,j}(x,y)$ is given by: 
\begin{align}\label{eq::pgfl::reduced::cluster::tcp}
{G}_{c_j}(v_{k,j}(x,{y})|{\bf z})= \exp\bigg( -\bar{m}_j\bigg(\int\limits_{0}^{\infty} (1-v_{k,j}({x},{y})) \Omega_j(y,z){\rm d}{y}\bigg)\bigg).
\end{align}
\end{cor}
\begin{IEEEproof}
Similar to Corollary~\ref{cor::pgfl::offspring::mcp},   the expression can be derived from Lemma~\ref{lemm::pgfl::cluster}  using Remark~\ref{rem::marginal::distribution}.
\end{IEEEproof}
For  PGFL of $\Phi_0$, i.e., $G_{0}(v_{0,k}(x,y))$, we  can  substitute  {$f_0(\cdot)$} in Lemma~\ref{lemm::pgfl_phi_0} with \eqref{eq::density_thomas_definition}. For \caseS~2, 
\begin{align*}
G_0(v_{k,0}({x},{y})) = \int\limits_{0}^{\infty}\frac{1}{1+\frac{P_0\beta_k}{P_k}\big(\frac{x}{y}\big)^{-\alpha}}\frac{y}{\sigma_0^2}\exp\bigg(-\frac{y^2}{2\sigma_0^2}\bigg){\rm d}y.
\end{align*}
For \caseS~3, 
\begin{align*}
G_0(v_{k,0}({ x},{y})) = \int\limits_{0}^{\infty}G_{c_0}(v_{k,0}(x,y)|z_0)\frac{z_0}{\sigma_0^2}\exp\bigg(-\frac{z_0^2}{2\sigma_0^2}\bigg){\rm d}z_0.
\end{align*}

We finally provide the expression of per-tier coverage probability for $k\in\ncalK\setminus\{0\}$. 
\begin{cor}\label{corr::per::tier::cov::tcp::k}
Per-tier coverage probability for $k\in\ncalK_2$ when all BS tiers in $\ncalK_2$ are modeled as TCPs can be expressed as:
\begin{align}\label{eq::per::tier::cov::tcp::k}
\pc_{k} = 2\pi\lambda_{{\rm p}_k}\bar{m}_k \int\limits_{0}^{\infty} \int\limits_{0}^{\infty}\prod\limits_{j\in\ncalK}G_j(v_{k,j}(x,y))G_{k}(v_{k,k}(x,y))G_{c_k}(v_{k,k}(x,y|{\bf z}))\Omega_k(x,z){\rm d}x\: z {\rm d}z.
\end{align}
\end{cor}
\begin{IEEEproof}
Similar to Corollary~\ref{corr::coverage::mcp},   the expression can be derived from Lemma~\ref{lemm::sumproduct::pcp} using Remark~\ref{rem::marginal::distribution}.
\end{IEEEproof}
We can obtain $\pc_0$ following the same arguments provided in the previous Section. 
\begin{cor}Per-tier coverage probability for $k=0$ when all BS tiers in $\ncalK_2$ are modeled as TCPs can be expressed as:
\begin{align*}
\pc_0 = \begin{cases}0,&\text{\caseS~1,}\\
\int\limits_{0}^{\infty}\prod\limits_{j\in\ncalK\setminus\{0\}}G_j(v_{0,j}(z_0,y))\bar{f}_0(z_0){\rm d}{z_0},&\text{\caseS~2},\\
\int\limits_{0}^{\infty}\int\limits_{0}^{\infty}\prod\limits_{j\in\ncalK\setminus\{0\}}G_j(v_{0,j}(x,y)) \exp\bigg(-\bar{m}_k \int\limits_{0}^{\infty} (1-v_{0,0}({ x},{ y})) \Omega_0(y,z_0) {\rm d}{ y}\bigg)\\
\times \Big(\bar{m}_0\int_{0}^{\infty} v_{0,0}(x, y)  \Omega_0(y,z_0) {\rm d}{ y} +1 \Big) \Omega_0({x},{z_0}){\rm d}x \: \frac{z_0}{\sigma_0^2}\exp\big(-\frac{z_0^2}{2\sigma_0^2}\big){\rm d}{z_0},&\text{\caseS~3}.
\end{cases}
\end{align*}
\end{cor}
\begin{IEEEproof}
Similar to Corollary~\ref{corr::pc_0_mcp},   
the expression can be derived from Lemma~\ref{lemm::sumproduct::finite} using Remark~\ref{rem::marginal::distribution}.
\end{IEEEproof}
\begin{figure}
 \begin{subfigure}{.48\textwidth}
  \includegraphics[width=\textwidth]{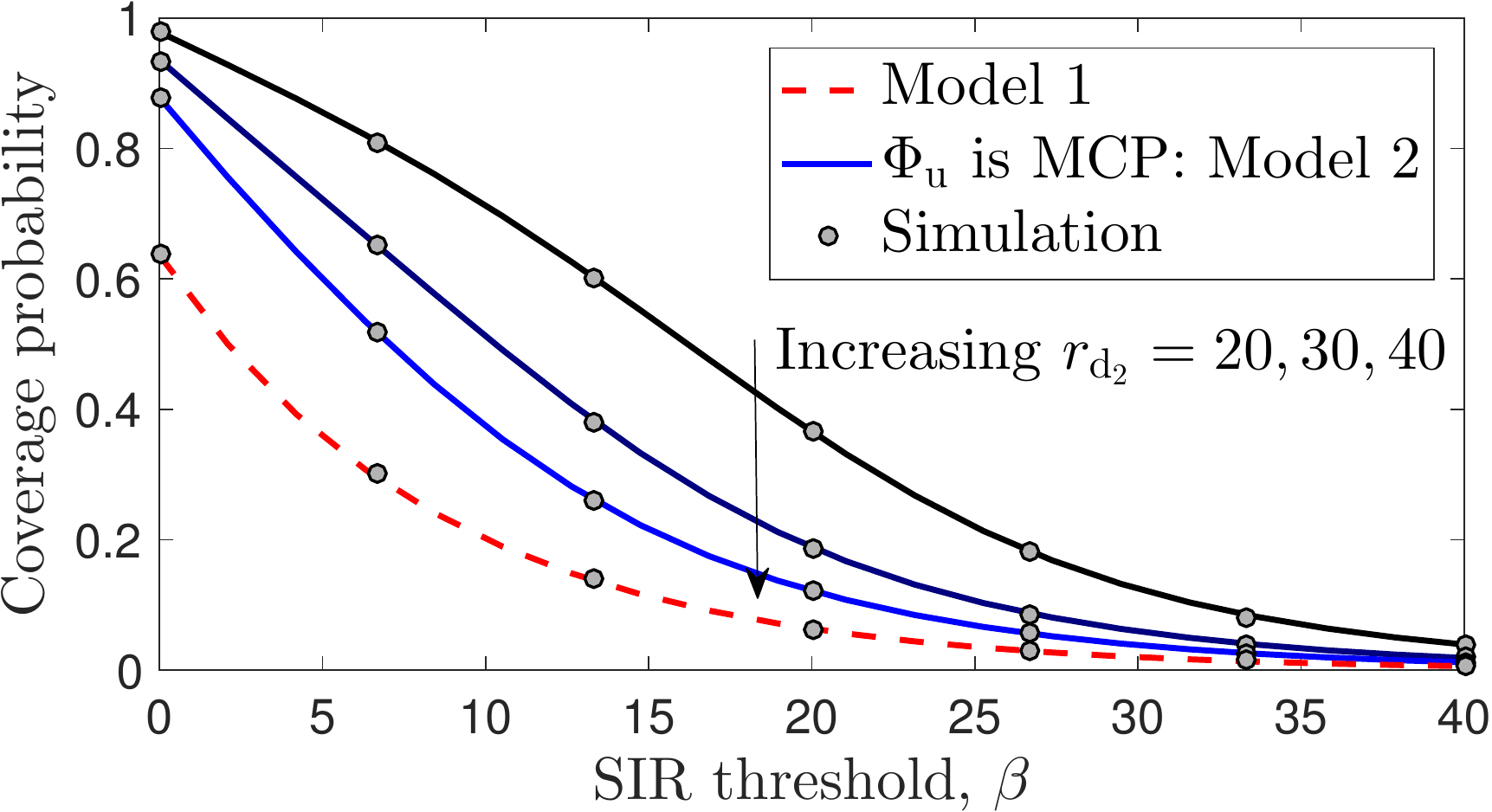}
\end{subfigure}%
\hfill
 \begin{subfigure}{.48\textwidth}
  \includegraphics[width=\textwidth]{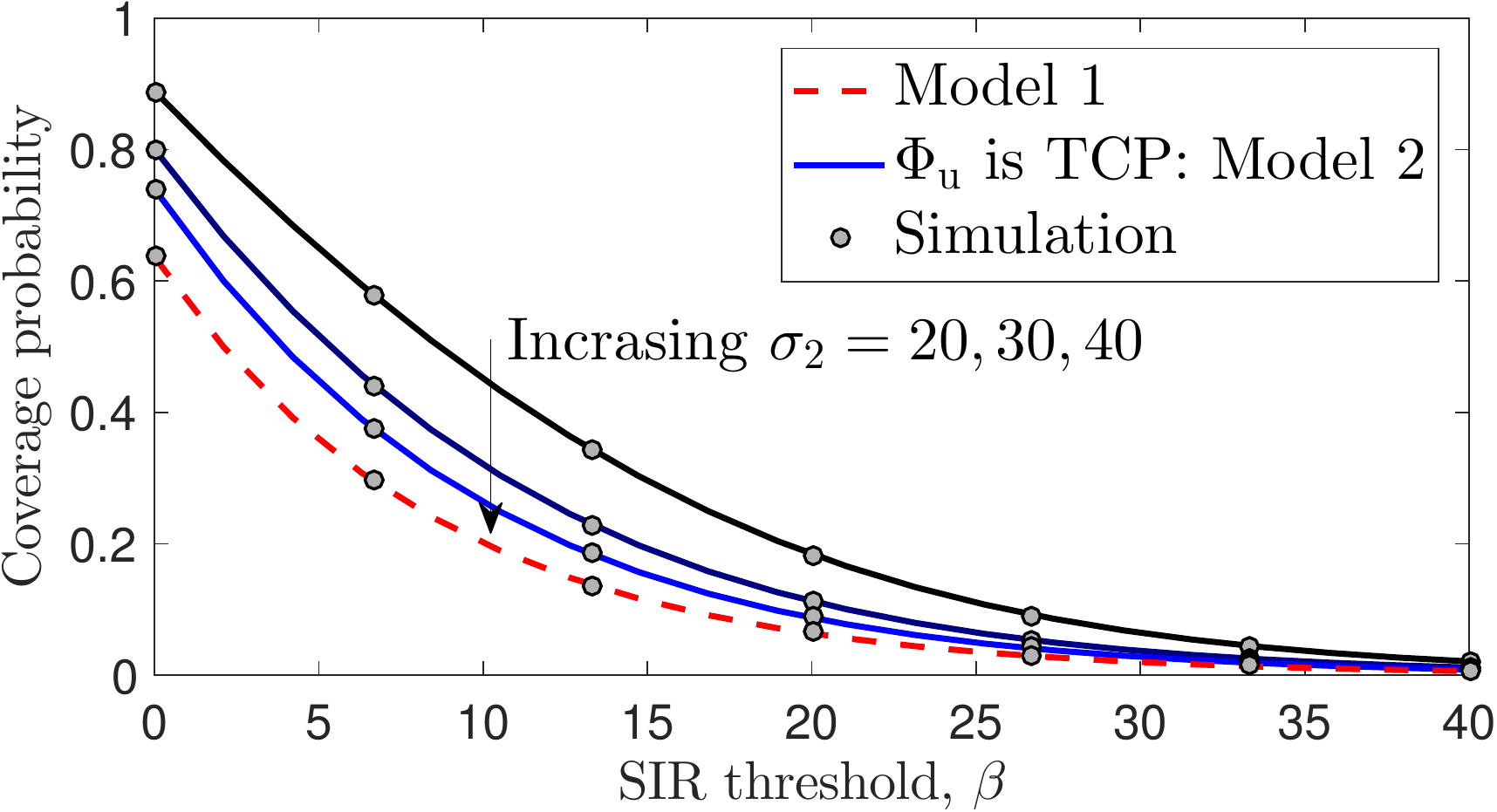}
\end{subfigure}
\caption{\small Coverage probability as a function of $\sir$ threshold ($\alpha=4$,  $\lambda_1=1$Km$^{-1}$,  $P_1=1000 P_2$, and $\lambda_2=100\lambda_1 $).}
\label{Fig:Model2}
\end{figure}
\begin{figure}
 \begin{subfigure}{.48\textwidth}
  \includegraphics[width=\textwidth]{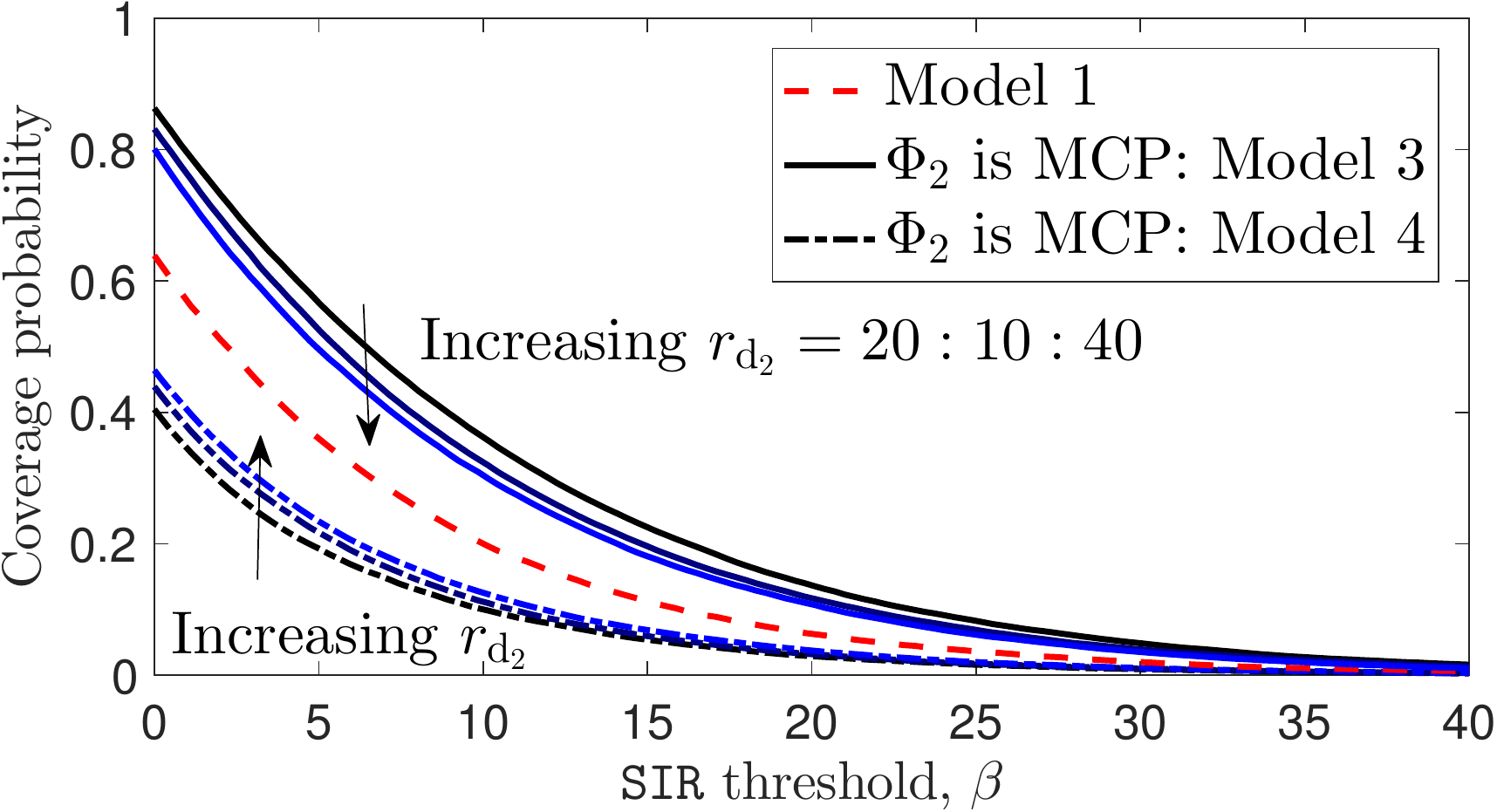}
\end{subfigure}%
\hfill
 \begin{subfigure}{.48\textwidth}
  \includegraphics[width=\textwidth]{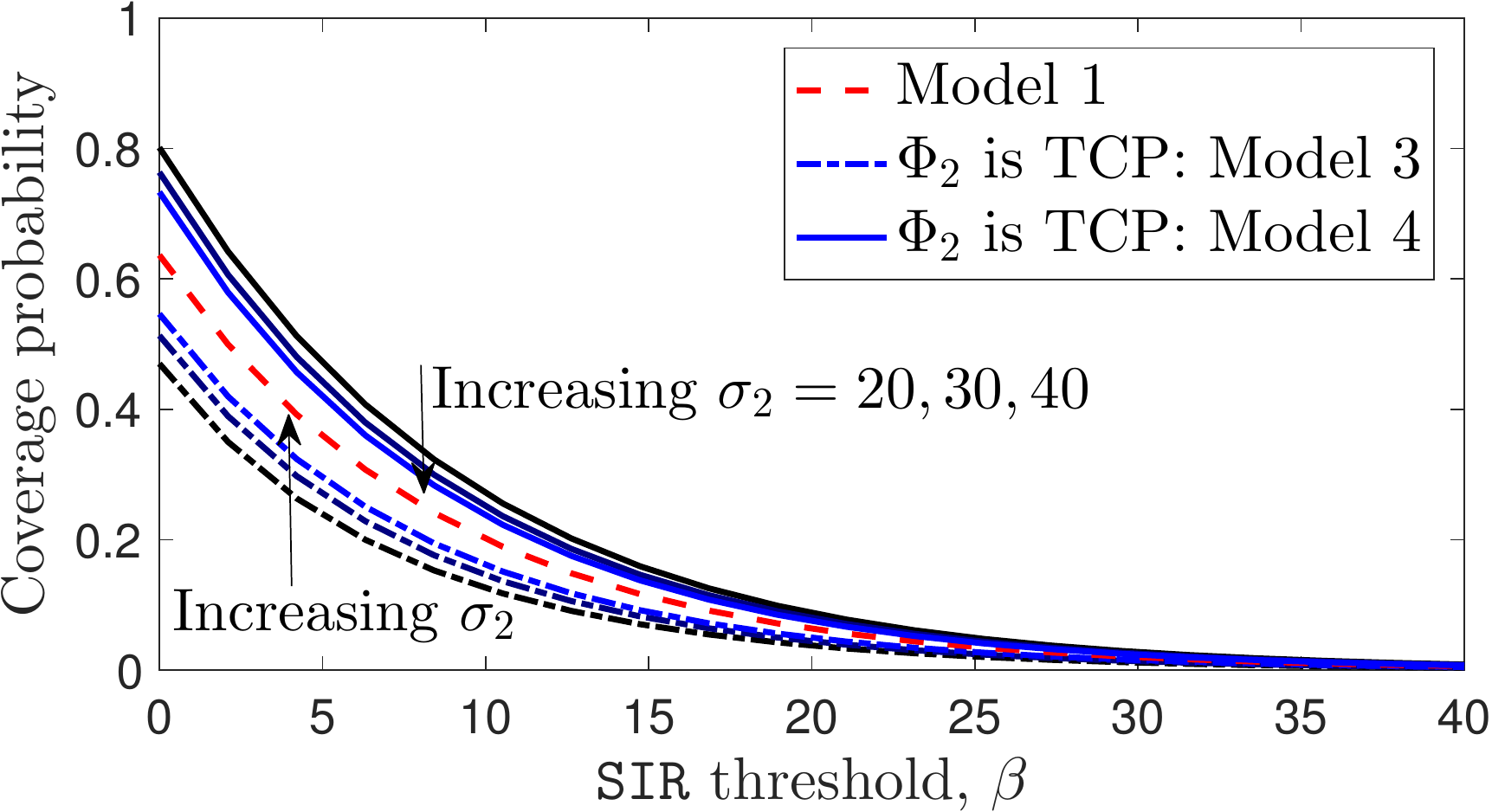}
\end{subfigure}
\caption{\small Coverage probability as a function of $\sir$ threshold ( $\lambda_1=1$Km$^{-1}$,  $P_1=1000 P_2$, and $\lambda_2=100\lambda_1 $).}
\label{Fig:Model34}
\end{figure}

\section{Results and Discussions}\label{sec::results}
In this Section, we compare the performance of  Models 1-4 introduced in Section~\ref{sec::hetnet_model} in terms of the coverage probability, $\pc$. 
 We first verify the analytical results with simulation of the $K$-tier HetNet. 
For all numerical results, we fix $\beta_1=\beta_2=\beta$, $\lambda_1=1\ {\rm Km}^{-1} $ and $\alpha =  4$.  All the BSs in the same tier transmit at fixed powers with $P_1/P_2 = 30\ {\rm dB}$. 
  For Models 1 and 2, we choose $K=2$, $\ncalK_1=\{1,2\}$, $\ncalK_2=\emptyset$. 
Users in Model 2 are distributed as a PCP, $\Phi_{\rm u}$ with $\Phi_2$ being the parent PPP. 
  For Models 3 and 4, we choose $K=2$, $\ncalK_1=\{1\}$, $\ncalK_2=\{2\}$. 
The perfect match between the simulation and analytical results verifies the accuracy of our analysis. 
From Figs.~\ref{Fig:Model2}-\ref{Fig:Model34}, we conclude that $\pc$ strongly depends on the choice of HetNet models. For instance, 
 a typical user experiences enhanced coverage in Model 2 than Model 1. From \figref{Fig:Model34}, we observe that $\pc$ of Model 1 is a lower bound on $\pc$ of Model 4 and is an upper bound on $\pc$ of Model 3. These observations bolster the importance of choosing appropriate models for different BS and user configurations that are cognizant of the coupling in the locations of the BSs and users. 
 

 \subsection{Effect of Variation of Cluster Size}
 We vary the cluster size of the PCP and observe the trend in $\pc$ for Models 2-4. For Model 2, we find in \figref{Fig:Model2} that $\pc$ decreases as cluster size (i.e. $r_{{\rm d}_2}$ for MCP, $\sigma_2$ for TCP) increases and   converges towards  that of  Model 1. 
 The reason of the coverage boost for denser cluster is that the SBS at cluster center lies closer to the typical user with high probability, hence improving the signal quality of the serving link. 
 Moving to Models 3 and 4 in \figref{Fig:Model34}, we again observe that $\pc$ of the two models converges to that of Model 1  as the cluster size (i.e. $r_{{\rm d}_2}$ for MCP, $\sigma_2$ for TCP) tends to infinity. We  proved this convergence in Section~\ref{sec::convergence}. We further observe from \figref{Fig:Model34}  that increasing cluster size has a conflicting effect on $\pc$ for Models~3 and~4: $\pc$ of Model~4 increases whereas that of  Model 3 decreases. This can be explained as follows. For Model 3, as cluster size increases,  the collocated user and SBS clusters become sparser and the candidate serving SBS lies farther to the typical user with high probability. 
On the contrary, for Model~4 where the user locations form an independent PPP,  the distance between the candidate serving SBS and the typical user decreases more likely with the increment of cluster size. 
\subsection{Effect of Variation of Intensity of Parent PPP}
\begin{figure}
 \begin{subfigure}{.48\textwidth}
  \includegraphics[width=\textwidth]{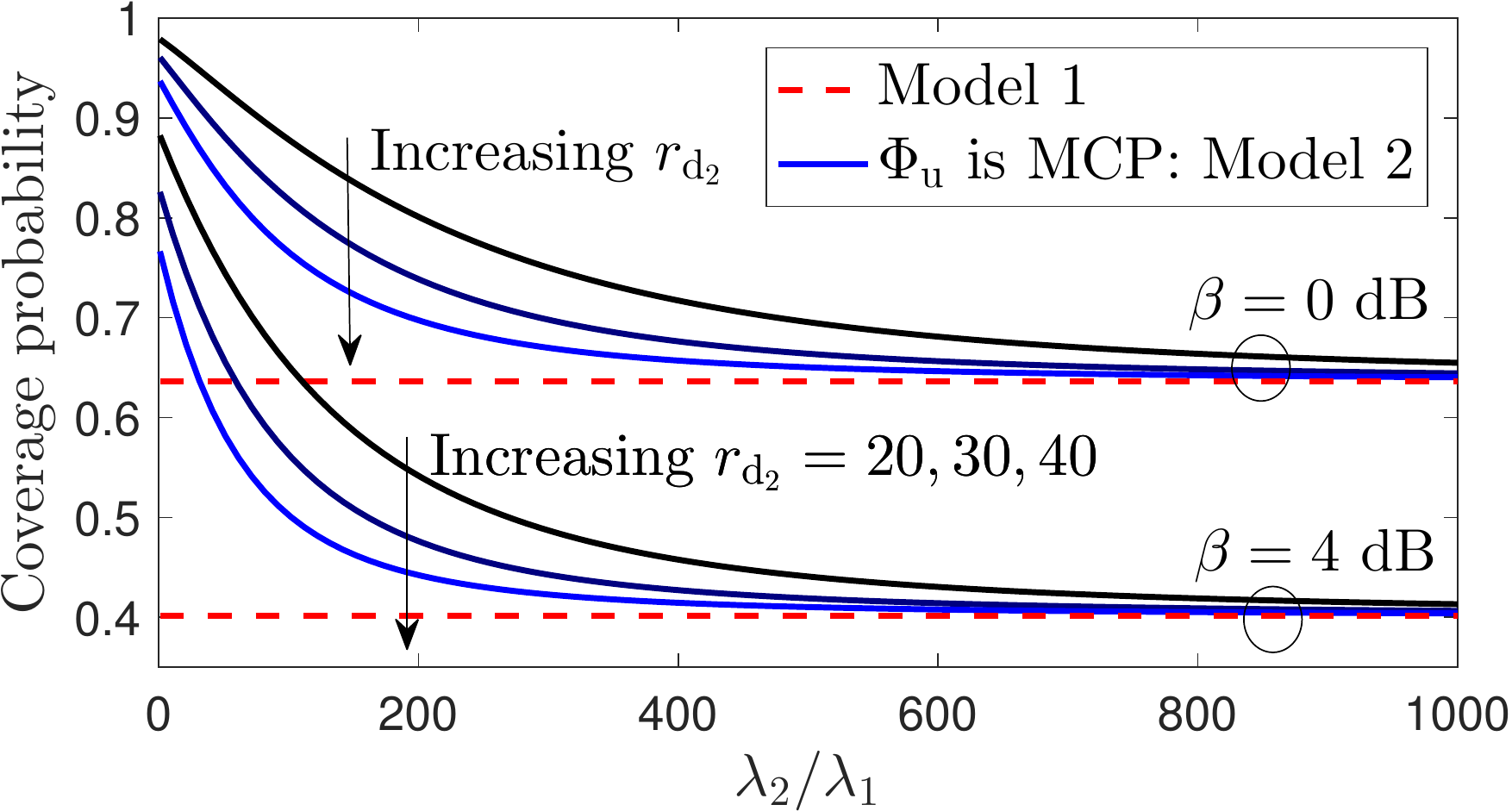}
\end{subfigure}%
\hfill
 \begin{subfigure}{.48\textwidth}
  \includegraphics[width=\textwidth]{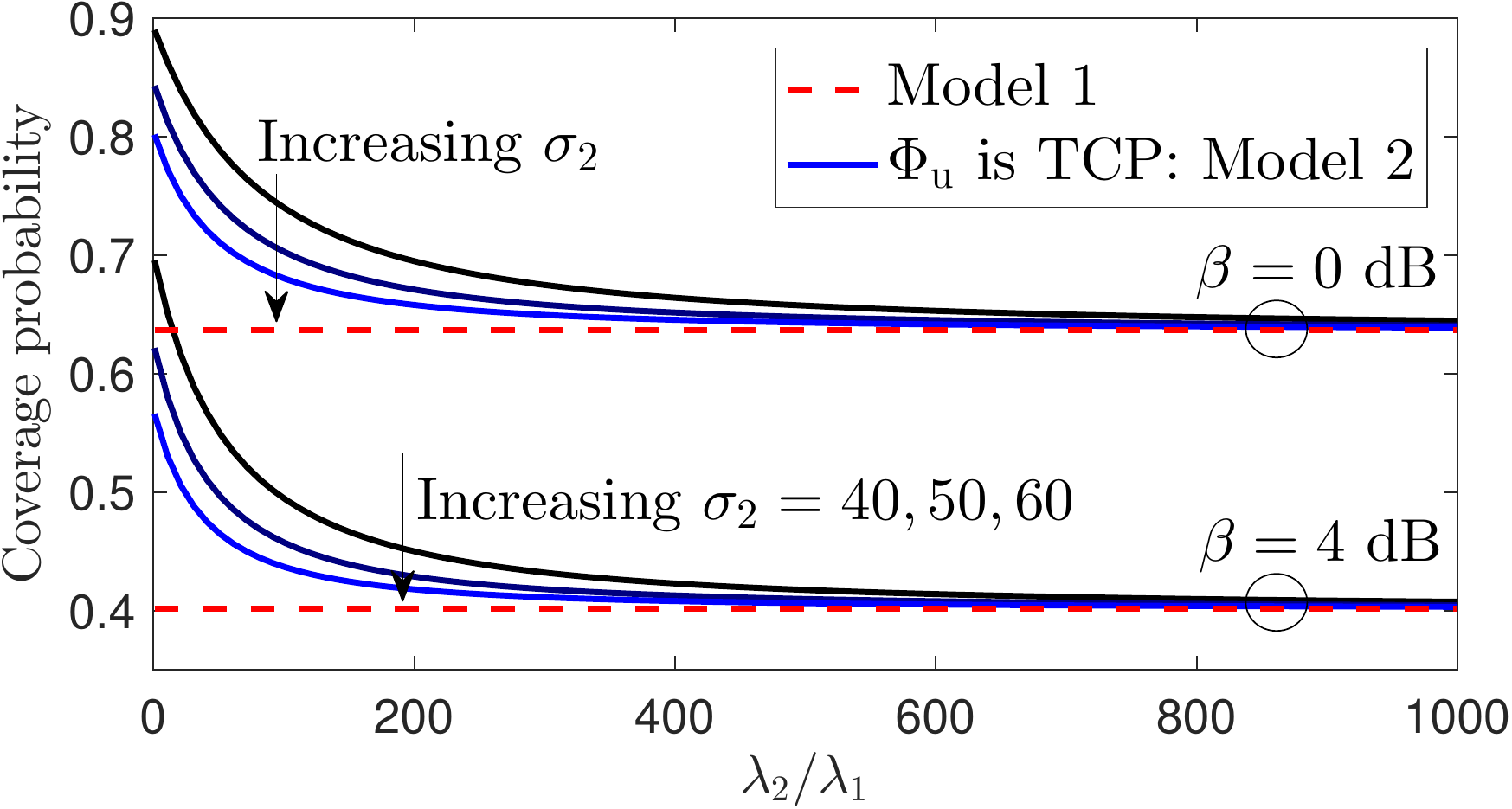}
\end{subfigure}
\caption{\small Coverage probability as a function of $\lambda_2/\lambda_1$ ($\lambda_1=1$Km$^{-1}$, and $P_1=1000 P_2$.).}
                \label{Fig:Model2::Lambda}
\end{figure}

\begin{figure}
 \begin{subfigure}{.48\textwidth}
  \includegraphics[width=\textwidth]{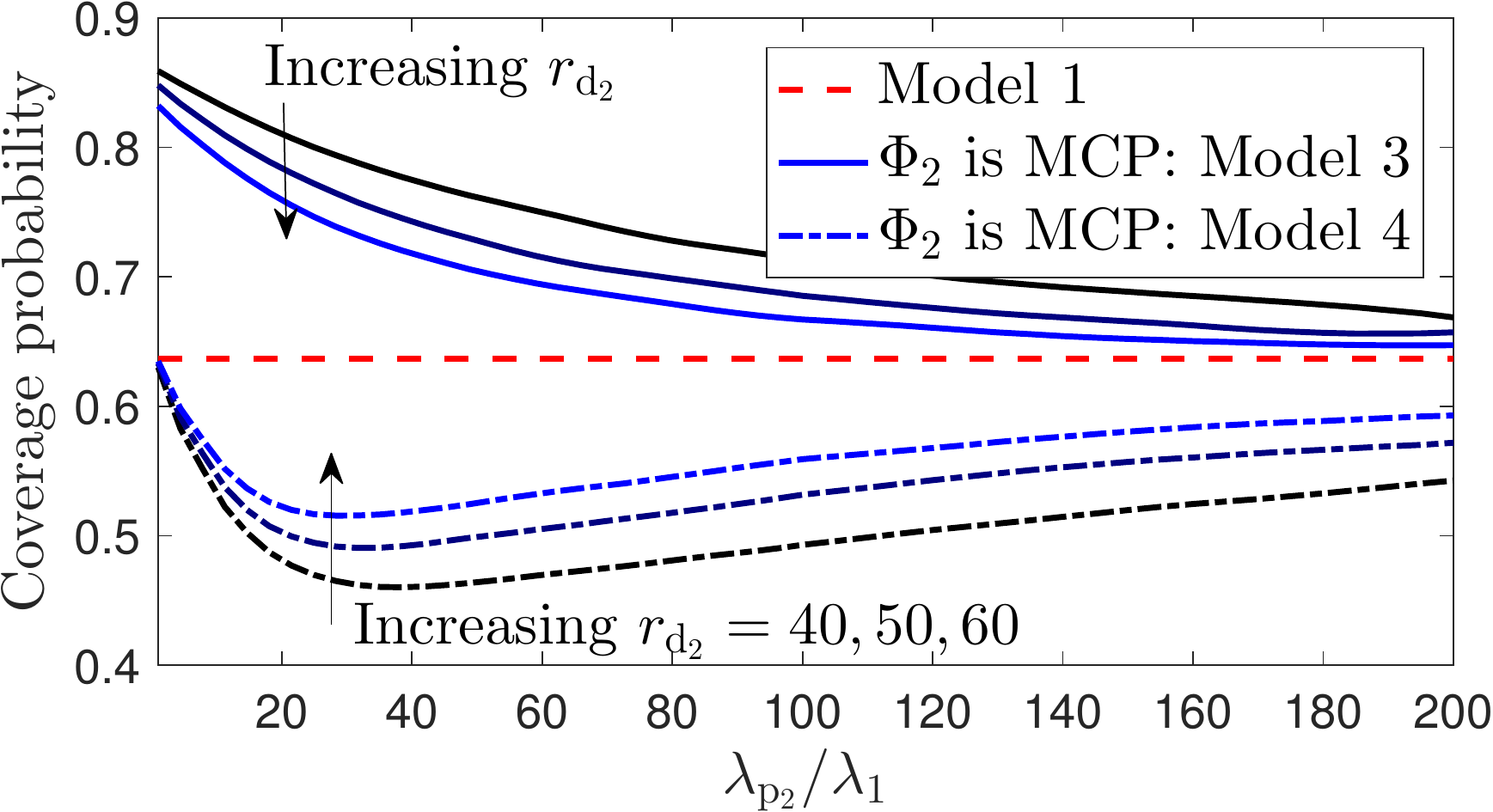}
\end{subfigure}%
\hfill
 \begin{subfigure}{.48\textwidth}
  \includegraphics[width=\textwidth]{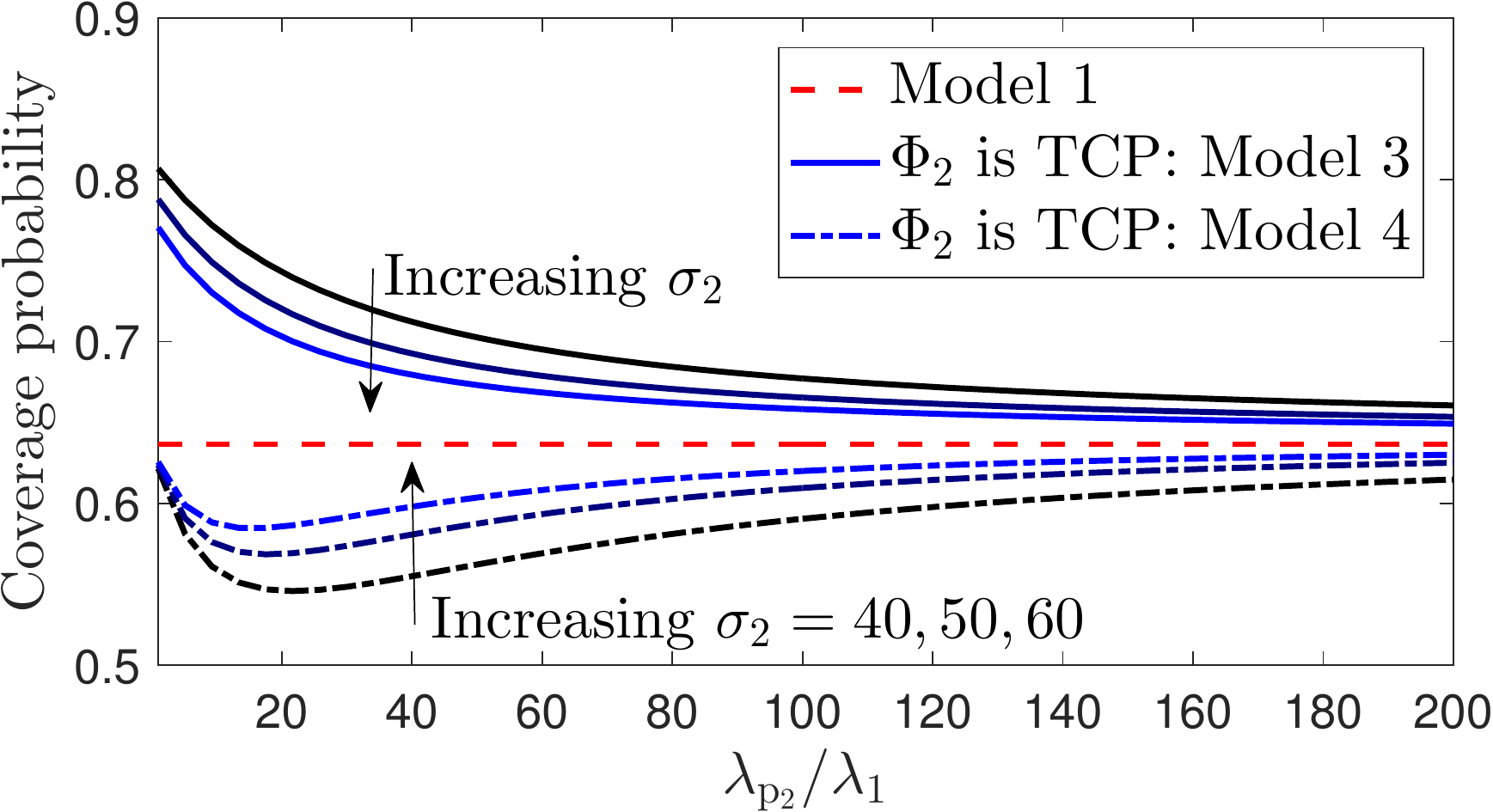}
\end{subfigure}
\caption{\small Coverage probability as a function of $\lambda_{{\rm p}_2}/\lambda_1$ ($\alpha=4$,  $\lambda_1=1$Km$^{-1}$, and $P_1=1000 P_2$.).}
                \label{Fig:Model34::Lambda}
\end{figure}

We study the effect of the variation of the intensity of the parent PPP  on $\pc$ for Models 2-4 ($\lambda_2$ for Models 2 and $\lambda_{{\rm p}_2}$, for Models 3 and 4) in Figs.~\ref{Fig:Model2::Lambda} and \ref{Fig:Model34::Lambda}. For Model 1, it is  well-known  that $\pc$ is independent of the intensities of BS PPPs \cite{dhillon2012modeling}. The intuition behind the observation is the fact that changing intensity of a PPP is equivalent to scaling the locations of all the points by  same factor. Hence the scaling factor cancels out from the serving and interfering powers in the $\sir$ expression. However, changing the intensity of the parent PPP of a PCP is not equivalent to the  location scaling of all the points by same factor. Thus, $\pc$ for Models 2-4 varies as a function of the intensity of the parent PPP. We also observe that as intensity of parent PPP increases, $\pc$ for Models 2-4 approaches to that of Model 1. 

\section{Conclusion}
In this paper, we developed a unified HetNet model by combining PPP and PCP that accurately models variety of spatial configurations for SBSs and users considered in practical design of HetNets, such as in the 3GPP simulation models. This is a significant generalization of the PPP-based $K$-tier HetNet model of \cite{5743604,dhillon2012modeling}, which was not rich enough to model non-uniformity and coupling across the locations of users and SBSs. For this model, we characterized the downlink coverage probability under max-$\sir$ cell association. As a part of our analysis, we evaluated the sum-product functional for PCP and the associated offspring point process. We also formally proved that a PCP weakly converges to a PPP when cluster size tends to infinity. Finally we specialized our coverage probability results assuming that the PCPs in the model are either TCPs or MCPs. 
This work has numerous extensions. An immediate extension is the coverage probability analysis with the relaxation of the assumption that the $\sir$-thresholds $\{\beta_k\}$ are greater than unity. From stochastic geometry perspective, this will necessitate the characterization of the $n$-fold Palm distribution \cite{coeurjolly2016tutorial,Keeler_factorial_moment} of PCP and its offspring point process. Extensions from the cellular network perspective involve analyzing other metrics like rate and spectral efficiency in order to obtain further insights into the network behavior. Coverage probability analysis under this setup for uplink is another promising future work. From modeling perspective, we can incorporate more realistic channel models, e.g. shadowing and general fading. 

\section*{Acknowledgment}
We would like to thank Prof. Jeffrey G. Andrews for helpful feedback. 
\appendix
\subsection{Proof of Theorem~\ref{thm::coverage}}\label{app::thm::coverage}
Under the assumption that $\beta_k>1,\ \forall\ k\in\ncalK$, there will be at most one BS in $\Phi$ satisfying the condition for coverage \cite{dhillon2012modeling}. 
Continuing from \eqref{eq::coverage_definition},
{
\begin{align}
&\pc=\sum\limits_{k\in\ncalK}\nbbE\bigg[\sum\limits_{{\bf x}\in\Phi_k}  {\bf 1}\bigg( \frac{P_k h_{\bf x}\|{\bf x}\|^{-\alpha}}{\ncalI(\Phi_k\setminus\{{\bf x}\})+\sum\limits_{j\in\ncalK\setminus\{k\}}\ncalI(\Phi_j)}>\beta_k\bigg)\bigg]\notag\\
&=\sum\limits_{k\in\ncalK}\nbbE\bigg[\sum\limits_{{\bf x}\in\Phi_k}\nbbP\big(h_{\bf x}>\frac{\beta_k}{{P_k}}\big(\ncalI(\Phi_k\setminus\{{\bf x}\})+\sum\limits_{j\in\ncalK\setminus\{k\}}\ncalI(\Phi_j)\big)\|{\bf x}\|^{\alpha}\big)\bigg]\notag\\
&\myeq{a}\sum\limits_{k\in\ncalK}\nbbE\bigg[\sum\limits_{{\bf x}\in\Phi_k}\exp\big(-\frac{\beta_k}{P_k}\big(\ncalI(\Phi_k\setminus\{{\bf x}\})+\sum\limits_{j\in\ncalK\setminus\{k\}}\ncalI(\Phi_j)\big)\|{\bf x}\|^{\alpha}\big)\bigg]\notag\\
&= \sum\limits_{k\in\ncalK}\nbbE\bigg[\sum\limits_{{\bf x}\in\Phi_k}\exp\bigg(-\frac{\beta_k}{P_k} \|{\bf x}\|^{\alpha}(\ncalI(\Phi_k\setminus\{{\bf x}\})\bigg)\Theta_k({\bf x})\bigg].\label{eq::coverage::intermediate}
\end{align}
}
Here, step (a) follows from $h_{\bf x}\sim \exp(1)$. The final step   follows from the independence  of $\Phi_k$, $\forall\ k\in\ncalK$, where,
\begin{align*}
&\Theta_k({\bf x})=\prod\limits_{j\in\ncalK\setminus\{k\}}\nbbE\exp\left(-\frac{\beta_k}{P_k}\|{\bf x}\|^{\alpha}
\ncalI(\Phi_j)\right)
=\prod\limits_{j\in\ncalK\setminus\{k\}}\nbbE\exp\left(-{\frac{\beta_k\|{\bf x}\|^{\alpha}}{P_k}\sum\limits_{{\bf y}\in\Phi_j} P_j{h_{\bf y}}\|{\bf y}\|^{-\alpha}}\right)\\
&=
\prod\limits_{j\in\ncalK\setminus\{k\}}\nbbE\prod\limits_{{\bf y}\in\Phi_j}\nbbE_{h_{\bf y}}\exp\left(-{\frac{\beta_k\|{\bf x}\|^{\alpha}}{P_k}{P_j h_{\bf y}}\|{\bf y}\|^{-\alpha}}\right)\myeq{a}  \prod\limits_{j\in\ncalK\setminus\{k\}}\nbbE\prod\limits_{{\bf y}\in\Phi_j}\frac{1}{1+ \beta_k \frac{ P_j}{P_k}\left(\frac{\|{\bf x}\|}{\|{\bf y}\|}\right)^{\alpha}}\\
&=\prod\limits_{j\in\ncalK\setminus\{k\}}G_j(v_{k,j}({\bf x},{\bf y})).
\end{align*}
Step (a) follows from the fact that $\{h_{\bf y}\}$ is an i.i.d. sequence of exponential random variables. 
Following from \eqref{eq::coverage::intermediate}, we get,
\begin{align*}
\pc&=  \sum\limits_{k\in\ncalK}\nbbE\bigg[\sum\limits_{{\bf x}\in\Phi_k}\Theta_k({\bf x})\exp\bigg(-\frac{\beta_k}{P_k} \|{\bf x}\|^{\alpha} \ncalI(\Phi_k\setminus\{{\bf x}\})\bigg)\bigg].
\end{align*}
The exponential term can be simplified following on similar lines as that of $\Theta_k({\bf x})$ and hence we obtain the final expression. 
{ \setstretch{1.2}
\bibliographystyle{IEEEtran}
\bibliography{Journal_ITA_arxiv2.bbl}
}
\end{document}